\documentclass{aa}

\usepackage{graphicx}
\usepackage{txfonts}
\begin{document}

   \title{A multi-wavelength study of Galactic H~{\sc II} regions with extended emission}

   \titlerunning{Galactic H~{\sc II} regions with extended emission}
   \authorrunning{Jyotirmoy Dey et al.}
   
   \author{Jyotirmoy Dey
          \inst{1}\fnmsep\thanks{Corresponding author}
          \and
          Jagadheep D. Pandian\inst{1}
          \and
          Dharam V. Lal\inst{2}
          \and
          Michael R. Rugel\inst{3}
          \and
          Andreas Brunthaler\inst{3}
          \and
          Karl M. Menten\inst{3}
          \and
          Friedrich Wyrowski\inst{3}
          \and
          Nirupam Roy\inst{4}
          \and
          Sergio A. Dzib\inst{3}
          \and
          Sac-Nict\'e X. Medina\inst{3}\fnmsep\inst{5}
          \and
          Sarwar Khan\inst{3}
          \and
          Rohit Dokara\inst{3}
          }

   \institute{Department of Earth and Space Sciences, Indian Institute of Space Science and Technology, Trivandrum, Kerala 695547, India            \email{dey.jyotirmoy04@gmail.com,jyotirmoydey.18@res.iist.ac.in}
        \and
            National Centre for Radio Astrophysics--Tata Institute of Fundamental Research, Post Box 3, Ganeshkhind P.O., Pune 411007, India
        \and
            Max Planck Institute for Radioastronomy (MPIfR), Auf dem H\"ugel 69, 53121 Bonn, Germany
        \and
            Department of Physics, Indian Institute of Science, Bengaluru 560012, India
        \and
            German Aerospace Center, Scientific Information, 51147 Cologne, Germany
             }

   \date{Received ; accepted }

% \abstract{}{}{}{}{} 
% 5 {} token are mandatory
 
    \abstract
   {H~{\small II} regions are the signposts of massive ($M\geq\,8\,M_\odot$) star-forming sites in our Galaxy. It has been observed that the ionizing photon rate inferred from the radio continuum emission of H~{\small II} regions is significantly lower ($\sim$ 90\%) than that inferred from far-infrared fluxes measured by the \textit{Infrared Astronomical Satellite}.} %Here, we study a sample of eight compact and ultracompact H~{\small II} regions to investigate this inconsistency.}
   {This discrepancy in the ionizing photon rates may arise due to there being significant amounts of dust within the H~{\small II} regions or the presence of extended emission that is undetected by high-resolution radio interferometric observations. Here, we study a sample of eight compact and ultracompact H~{\small II} regions with extended emission to explore its role in resolving the discrepancy.}
   %Thus, we picked our sample of H~{\small II} regions with extended emission to explore whether it could resolve the inconsistency.}
   {We have used observations at the \textit{upgraded Giant Metre Wave Radio Telescope} (1.25--1.45~GHz) and data from the GLOSTAR survey (4--8~GHz) to estimate the ionizing photon rate from the radio continuum emission. 
   %These observations have high angular resolution and good sensitivity to the diffuse emission and are hence well suited to investigate the role of dust and extended emission in both compact and ultracompact H~{\small II} regions.
   We have also estimated the ionizing photon rate from the infrared luminosity by fitting a spectral energy distribution function to the infrared data from the \textit{Spitzer}-GLIMPSE, MIPSGAL, and Hi-GAL surveys. The excellent sensitivity of the radio observations to extended emission allows us to investigate the actual fraction of ionizing photons that are absorbed by dust in compact and ultracompact H~{\small II} regions.}
   {Barring one source, we find a direct association between the radio continuum emission from the compact and diffuse components of the H~{\small II} region.
   %that in all cases, the extended emission is smoothly connected to the emission from the dense components.
   Our study shows that the ionizing photon rates estimated using the radio and infrared data are within reasonable agreement (5--28\%) if we include the extended emission. We also find multiple candidate ionizing stars in all our sources, and the ionizing photon rates from the radio observations and candidate stars are in reasonable agreement.}
   {}


   \maketitle
%
%-------------------------------------------------------------------

\section{Introduction}\label{sec:intro}

Massive stars form copious amounts of Lyman-continuum photons (Ly-photon hereafter) that ionize their surroundings, creating H~{\small II} regions. Initially, dense (sizes~$\lesssim 0.03$~pc, density~$\gtrsim 10^6$~cm$^{-3}$, and emission measure $\gtrsim 10^{10}$~pc~cm$^{-6}$; \citealt{2005IAUS..227..111K}) regions of ionized gas, known as hypercompact H~{\small II} regions, form around the O and B-type stars, which evolve into ultracompact~H~{\small II} regions (UCHRs; sizes~$\lesssim 0.1$~pc, density~$\gtrsim 10^4$~cm$^{-3}$, and emission measure $\gtrsim 10^7$~pc~cm$^{-6}$; \citealt{1989ApJS...69..831W,1994ApJS...91..659K,2005IAUS..227..111K}). The UCHRs continue to expand with time, eventually leading to the formation of compact and classical H~{\small II} regions. Although the H~{\small II} regions are generally bright at radio and infrared wavelengths, most of them are invisible at optical wavelengths because of the high degree of obscuration from the surrounding dust. The first studies of H~{\small II} regions were carried out by \cite{1966ApJ...144..937D}, \cite{1967ApJ...147..471M}, and \cite{1967ApJ...148L..17R} using single-dish radio telescopes to understand the nature of their continuum emission. The first large-scale interferometric survey of UCHRs was conducted by \cite{1989ApJS...69..831W} (WC89 hereafter). They noticed that the Ly-photon rate estimated from the \textit{Infrared Astronomical Satellite} (IRAS) fluxes are significantly higher than that obtained from the radio continuum. A similar disagreement was found by \cite{1994ApJS...91..659K}, with the discrepancy being attributed to the absorption of the ionizing photons by the dust enclosed within the ionized region.

\citet{1994ApJS...91..659K} also suggested an alternate explanation invoking the presence of a stellar cluster to explain the discrepancy above. Assuming that the mass distribution of stars in the cluster follows the typical initial mass function (e.g., \citealt{2001MNRAS.322..231K}), a massive star will be accompanied by a number of lower mass stars which will contribute to the infrared luminosity but not to the rate of ionizing photons. However, adopting an initial mass function from \citet{2001MNRAS.322..231K} and the mass-luminosity relation for main sequence stars \citep{2005essp.book.....S}, the massive stars are found to account for greater than 90\% of the total luminosity. Hence, the presence of a stellar cluster does not significantly alter the statistics of the discrepancy between the rate of ionizing photons inferred from radio and infrared wavelengths.

Meanwhile, later studies of \cite{1999ApJ...514..232K}, \cite{2001ApJ...549..979K}, \cite{2005MNRAS.357.1003E}, and \cite{2020MNRAS.492..895D} showed that the presence of extended radio continuum emission that is directly associated with the UCHRs might also resolve the inconsistency between the Ly-photon rates. The extended emission is often undetected due to the poor sensitivity of interferometric observations to the emission on scales larger than that corresponding to the shortest baseline of the interferometer. The much larger radio continuum flux from the inclusion of extended emission gives rise to larger Ly-photon rates that are closer to those inferred from the infrared, thereby avoiding the need for high dust absorption as suggested by WC89 and \cite{1994ApJS...91..659K}. A similar scenario may also occur for the compact H~{\small II} regions, where the dust absorption could be overestimated due to the non-detection of extended emission surrounding the dense, compact cores \citep{habing1979compact}. Another problem that may be addressed by including the extended emission is the ``age problem'' wherein the H~{\small II} regions must remain in the UC phase for much longer than their sound crossing time in a homogeneous medium. The presence of extended emission surrounding the UCHRs will suggest that those regions are not as young as they appear from the observation of the UCHR alone. 

%\textbf{In addition to a physically associated extended emission, a few additional solutions are also proposed to solve this inequality. \citet{1994ApJS...91..659K} hinted that the presence of a cluster with a most massive member of mass $M_{\text{u}}$ and a distribution of lower mass stars could also produce the observed infrared luminosity, while not contributing much towards the ionization of the gas inferred from the radio observations. Comparing radio with infrared cluster spectral types, the authors found that the values generally differed by no more than one and a half sub-classes. Hence, they assumed that the significantly high Ly-photon rate (estimated from the infrared measurements) could be an overestimate due to the single-star assumption.}

%\textbf{However, while testing this theory, assuming a stellar cluster as specified by typical initial mass functions \citep{2001MNRAS.322..231K}, for a few of the WC89 sources (assuming the mass of the single ZAMS star as the upper limit of mass), this solution ceases to resolve the inconsistencies between the Ly-photon rates. We also found that a physically associated extended emission contributes much more towards resolving this inconsistency. Hence, we focussed our study on detecting an extended emission surrounding the compact and ultracompact cores.}

In this paper, we present results from a study of extended emission around eight compact and ultracompact Galactic H~{\small II} regions using the \textit{upgraded Giant Metrewave Radio Telescope} (uGMRT; \citealt{1990IJRSP..19..493S, gupta2017upgraded}) data covering 1.25--1.45~GHz, and 4--8~GHz data from the GLOSTAR (\textit{A global view on star formation}) survey (\citealt{2019A&A...627A.175M,2021A&A...651A..85B}). Both the uGMRT and GLOSTAR data are ideally suited to address the presence or absence of extended emission around compact and ultracompact H~{\small II} regions. The uGMRT comprises a hybrid configuration of 30 antennas (shaped like a `Y') with a central array consisting of 14 antennas within a 1 sq km region and the remaining 16 antennas on the 3 arms of the `Y'. This gives excellent sensitivity to extended emission on scales smaller than 7$\arcmin$ along with good angular resolution ($\approx2\arcsec$ at 1.4~GHz). The GLOSTAR survey comprises observations with the \textit{Karl G. Jansky Very Large Array} (VLA) in its D and B-configurations along with zero spacing information from the 100-m \textit{Effelsberg radio telescope}. The sensitivity of both uGMRT and GLOSTAR observations to extended emission allows us to carry out a comprehensive study of the role of the same in resolving outstanding questions of Galactic H~{\small II} regions.
%Both the uGMRT observations and GLOSTAR data were able to detect extended emission in both radio continuum and radio recombination lines (RRLs), from which we can carry out a comprehensive study of both statistics about Ly-photon rates and gas dynamics in the Galactic H~{\small II} regions. 

Although extended emission is frequently detected towards H~{\small II} regions, it may or may not be associated with the compact or ultracompact emission. \cite{1999ApJ...514..232K} proposed two methods to determine whether or not an extended emission is associated with compact or ultracompact cores. The first is to search for a discontinuity in the intensity of the continuum emission (a gap with near-zero emission). The presence of such a gap between the core and extended components would suggest that the two regions are physically distinct. However, one has to be careful applying this technique to interferometer data since a discontinuity can be created due to the filtering of large-scale emission. An alternate method is to look at the kinematics of the ionized gas using radio recombination lines (RRLs). The presence or absence of a smooth velocity distribution between the compact core and extended emission can be used to infer whether or not the two structures are physically associated. For example, \citet{2001ApJ...549..979K} used H76$\alpha$ observations to examine (using the RRL velocity field) whether the compact and extended emission are physically connected in 16 H~{\small II} regions and found physical association in all sources with one exception.

Our observations include RRLs in addition to radio continuum to establish the association of emission from the compact and extended components. Further, previous studies by \citet{1972MNRAS.157..179B,1975ApJ...201..134L,1978ARA&A..16..445B} have shown that low-density ($n_{\text{e}} \approx 10$~cm$^{-3}$) ionized gas emits more low-frequency or high-n (Hn$\alpha$; n $\approx$ 150--180) RRLs compared to intermediate- or high-density gas ($n_{\text{e}} \approx 10^2$~cm$^{-3}$ and $n_{\text{e}} \approx 10^3$~cm$^{-3}$), which show enhanced emission of low-n (n $\approx$ 80--120) RRLs (Fig.~1 of \citealt{1978ARA&A..16..445B}). Since our uGMRT observations cover the H166--172$\alpha$ lines, and the GLOSTAR survey observes the H98--114$\alpha$ lines, we expect to be able to trace line emission from ionized gas over a range of densities, including the compact and extended regions. 

This paper is organized as follows. We describe the source selection criteria in Sect.~\ref{sec:srcs}. The radio observations with the uGMRT and subsequent data analysis using observed as well as archival data are described in Sect.~\ref{sec:obs}. Sect.~\ref{sec:results} presents the results obtained with notes on individual sources described in Sect.~\ref{sec:notes}. In Sect.~\ref{sec:discuss}, we discuss the association of the extended emission with the H~{\small II} regions and its role in resolving the outstanding problems associated with them.

%--------------------------------------------------------------------

\section{Source selection} \label{sec:srcs}

Since the primary focus of our study is to observe H~{\small II} regions whose morphology at radio wavelengths comprises extended emission surrounding compact and ultracompact cores, we selected candidate sources that have an extended morphology visible from radio to mid-infrared wavelengths. First, we looked at sources classified as Galactic H~{\small II} regions in the catalog of \textit{The H~{\small I}, OH, recombination line survey of the Milky Way} (THOR; \citealt{2016A&A...595A..32B, 2018yCat..36190124W}). Considering the availability of calibrated images from the GLOSTAR survey at the time of the uGMRT observations, we constrained ourselves to the Galactic longitude region $18^{\circ} \leq l \leq 30^{\circ}$. The sources were selected to have a size of at least $1'$ so that the extended emission would be well detected in both the uGMRT observations and the GLOSTAR survey. Since one of our objectives is to study the dynamics of ionized gas, we also ensured that all sources have detections of RRLs in both THOR and GLOSTAR surveys. Finally, we examined the field surrounding the targets and only selected sources that are devoid of nearby bright sources, as the presence of such sources affects data calibration by contamination through the sidelobes of the primary beam. These selection criteria provided us with a sample of eight Galactic H~{\small II} region candidates to observe using uGMRT (see Table~\ref{tab:source}). The kinematic distances to the sources range from 1.8 to 11.7~kpc (adopted from \citealt{2018MNRAS.473.1059U} and references therein) while the angular size ranges from 1.5$'$ to 3.7$'$.

It is to be noted that the angular sizes in Table~\ref{tab:source} translate to physical sizes ranging from 0.9 to 9.2~pc, which are much larger than those of UCHRs. This is due to the detection of extended emission by the THOR survey. The nature of the target H~{\small II} regions, including whether or not the sources are bona fide UCHRs, is examined in Sect.~\ref{sec:radioresults}.

\begin{table*}
\caption{Properties of the target sources.
\label{tab:source}}
\centering
\begin{tabular}{c c c c c c c c}
\hline\hline 
Source name & RA (J2000)$^\text{a}$ & DEC (J2000)$^\text{a}$ & d$^\text{b}$ & $V_\text{LSR}$$^\text{b,c}$ & $D$ & $S_{\text{p}}$ & $S_{\text{int}}$ \\
 & (hms) & (dms)  & (kpc)  & (km~s$^{-1}$) & (arcmin) & (Jy~beam$^{-1}$) & (Jy) \\
\hline

$^{*}$G19.68$-$0.13 & 18$^\text{h}$27$^\text{m}$23.40$^\text{s}$ & --11$\degr$49$\arcmin$52.78$\arcsec$ & 11.7 & 55.0 & 1.5 & 0.211 & $0.683 \pm 0.015$  \\
$^{*}$G20.99$+$0.09 & 18$^\text{h}$29$^\text{m}$04.73$^\text{s}$ & --10$\degr$34$\arcmin$15.59$\arcsec$ & 1.8 & 18.6 & 1.8 & 0.228 & $1.221 \pm 0.011$ \\
$\;\,$G22.76$-$0.48 & 18$^\text{h}$34$^\text{m}$26.70$^\text{s}$ & --09$\degr$15$\arcmin$48.66$\arcsec$ & 4.9 & 74.8 & 3.6 & 0.164 & $1.703 \pm 0.014$  \\
$\;\,$G24.47$+$0.49 & 18$^\text{h}$34$^\text{m}$10.19$^\text{s}$ & --07$\degr$17$\arcmin$57.96$\arcsec$ & 5.9 & 99.8 & 2.1 & 0.798 & $3.452 \pm 0.022$  \\
$\;\,$G25.69$+$0.03 & 18$^\text{h}$38$^\text{m}$04.03$^\text{s}$ & --06$\degr$25$\arcmin$31.45$\arcsec$ & 11.7$^{\text{d}}$ & 51.9 & 2.7 & 0.149 & $0.961 \pm 0.017$  \\
$^{*}$G27.49$+$0.19 & 18$^\text{h}$40$^\text{m}$49.01$^\text{s}$ & --04$\degr$45$\arcmin$04.04$\arcsec$ & 2.2 & 32.5 & 3.7 & 0.252 & $1.873 \pm 0.020$ \\
$\;\,$G28.30$-$0.39 & 18$^\text{h}$44$^\text{m}$21.85$^\text{s}$ & --04$\degr$17$\arcmin$37.16$\arcsec$ & 9.7 & 85.5 & 1.5 & 0.203 & $0.746 \pm 0.019$ \\
$^{*}$G28.81$+$0.17 & 18$^\text{h}$43$^\text{m}$16.84$^\text{s}$ & --03$\degr$35$\arcmin$27.23$\arcsec$ & 7.9 & 103.0 & 2.1 & 0.358 & $1.605 \pm 0.016$ \\
\hline
\end{tabular}
\tablefoot{ \tablefoottext{*}{sources comprise of multiple fragments of compact emission,} \tablefoottext{a}{coordinates of the H~{\small II} region from the radio continuum catalog of \citet{2018yCat..36190124W},} \tablefoottext{b}{adopted kinematic distance from \citet{2018MNRAS.473.1059U} and references therein,} \tablefoottext{c}{LSR velocities from \citet{1989ApJS...71..469L},} \tablefoottext{d}{revised to far distance based on the H~{\small I} absorption spectrum,} $D$ = size of the UCHR (including the extended emission) using THOR data, $S_{\text{p}}$ = peak intensity at the 1.31~GHz spectral window of the THOR survey, $S_{\text{int}}$ = integrated intensity at the 1.31~GHz spectral window of the THOR survey.}
\end{table*}

%--------------------------------------------------------------------

\section{Observations and data analysis} \label{sec:obs}

\subsection{uGMRT observations and data reduction}

We observed the targets using the \textit{upgraded Giant Metrewave Radio Telescope} (uGMRT; \citealt{1990IJRSP..19..493S, gupta2017upgraded}) with the GWB correlator as the backend (proposal codes 37\_073 and 40\_100; see Table~\ref{tab:table1} for the dates of observations). The correlator was configured to have a bandwidth of 200~MHz centered at 1350~MHz (Band-5) with 8192 channels. A total of seven RRLs (H166$\alpha$ -- H172$\alpha$) were covered in this bandwidth. At the observed frequency, uGMRT has a native resolution of $\approx 2\arcsec$ and the largest detectable angular scale of $\approx 7\arcmin$ in Band-5. The radio sources 3C48 and 3C286 were used as the bandpass and flux calibrators according to their availability, whereas J1822$-$096 was used as the gain calibrator, which has a flux density of 5.6~Jy in this band. The on-source time ranged from 2 to 8~hours depending on their peak and extended flux densities, as seen in the THOR survey.

\begin{table}
\caption{Details of our observation using uGMRT.}
\label{tab:table1}
\begin{tabular}{lc}
\hline\hline
Parameter & Value \\
\hline
		Observation date & 2019 Dec 1, 2, 6, 10, 14 \\
		                  & 2021 Sep 14, 26, 27, 28 \\
                    & 2021 Oct 5, and 2021 Nov 28 \\
%		System Temperature & 73 K \\
%		Time allotted & 62 hrs\\
		No. of channels & 8192\\
		Central frequency & 1350 MHz\\
		Bandwidth & 200 MHz \\
		Primary Beam  & $25\arcmin$ \\
		Synthesised Beam & $\approx2\arcsec$ \\
\hline
\end{tabular}
\end{table}

The data were analyzed using the NRAO \textit{Common Astronomy Software Applications} (\texttt{CASA}; \citealt{2007ASPC..376..127M}) package. The bad data were flagged using the \texttt{flagdata} task. Next, gain solutions were computed and applied to the calibrators. Then, the data examination and flagging were carried out for another round, and the final bandpass and gain solutions were computed. Next, the solutions were applied to the target source.

Although the channels affected by Radio Frequency Interference (RFI) were flagged before calibration, another careful examination of the data was performed before imaging the targets. After completing this step, the net bandwidth left for imaging the targets was around 150~MHz, considering only the line-free channels. The targets were then imaged using the \texttt{tclean} task, and the dynamic range of the continuum images was improved by self-calibration. The final 1$\sigma$ noise in the continuum images is 60--70~$\mu$Jy~beam$^{-1}$, and the synthesized beam ranged from 1.7--3.2$''$. The uncertainty in flux calibration is estimated to be better than 10\%, based on comparing the flux density of the gain calibrator to its standard value after applying the gain solutions. The flux densities of the target sources were also compared with that of the nearest sub-band of the THOR survey and were found to agree to better than 10\%.

As shown in Table~\ref{tab:table1}, 8192 channels were used to observe the targets, providing a native velocity resolution of $\sim 5$~km~s$^{-1}$. To image the RRLs, we first subtracted the continuum emission using the task \texttt{UVLSF} of the NRAO \textit{Astronomical Image Processing System} (\texttt{AIPS}). We used \texttt{UVLSF} of \texttt{AIPS} for this purpose as opposed to the \texttt{uvcontsub} task of \texttt{CASA} since the former achieved better continuum subtraction given the complex shape of the spectral baseline after bandpass calibration. Before imaging the RRLs, we applied the self-calibration solutions determined from the radio continuum to the line data.
 
The RRL data were imaged using the \texttt{clean} task. Since the RRLs are weak, especially in the regions of extended emission, we used a restoring beam of 25$''$ for imaging the RRLs. To further enhance the signal-to-noise ratio, four RRL images (H167$\alpha$ -- H170$\alpha$; the other lines were either affected by RFI or were located at the edges of the passband) were stacked to the observed frequency of H169$\alpha$ using the package \texttt{LineStacker} \citep{2020MNRAS.499.3992J}\footnote{\url{https://github.com/jbjolly/LineStacker/releases}}. This results in an improvement in the signal-to-noise ratio of the line image by a factor of 2, with the final images having an rms noise of $\sim 0.30$~mJy~beam$^{-1}$ per 5 km~s$^{-1}$ spectral channel.

\subsection{Data from the GLOSTAR survey}

The uGMRT observations are complemented by data from the GLOSTAR survey \citep{2019A&A...627A.175M, 2021A&A...651A..85B}. The GLOSTAR survey was carried out with the VLA in its D and B-configurations, using a correlator configuration that covers the continuum emission in full polarization within 4--8~GHz (using two 1~GHz sub-bands centered at 4.7 and 6.9~GHz) in addition to the 6.7~GHz methanol maser line, the 4.8~GHz formaldehyde line, and seven RRLs. The zero spacing information was covered using the 100-m \textit{Effelsberg radio telescope}.

The data products from the survey include images from the D-configuration only (GLOSTAR-D hereafter; \citealt{2019A&A...627A.175M}), B-configuration only (GLOSTAR-B hereafter; \citealt{2023A&A...670A...9D,2023A&A...680A..92Y}), combined B and D-configurations, and D-configuration combined with Effelsberg (GLOSTAR-D+Eff hereafter; \citealt{2023A&A...671A.145D}). The continuum data products from GLOSTAR-D and GLOSTAR-B include a single continuum image at an effective frequency of 5.79~GHz and eight sub-band images that can be used to measure the spectral index. Since the angular resolution has a dependence on both frequency as well as location in the plane of the sky, all continuum images have a circular restoring beam of 18$\arcsec$ and 1$\arcsec$ for the D and B-configurations, respectively. The largest detectable structures in the GLOSTAR-D data is $\sim 2\arcmin$, while that in GLOSTAR-B is $4''$ due to the imaging being carried out using only data with $uv$-ranges greater than 50~k$\lambda$. The GLOSTAR-D+Eff continuum image has an effective frequency of 5.85~GHz and an angular resolution of 18$\arcsec$. Due to their weak strength, the RRLs are only imaged in the D-configuration \citep{2024khan}, and all detected RRLs are stacked to increase the signal-to-noise ratio. The stacked RRL map has a velocity resolution of 5~km~s$^{-1}$ and an angular resolution of 25$\arcsec$.

\subsection{FIR emission from cold dust}\label{analysis:fir}

We have used data from the Hi-GAL survey \citep{2010A&A...518L.100M} to construct the spectral energy distribution (SED) of dust emission from mid- to far-infrared wavelengths. The Hi-GAL survey provides images of the far-infrared continuum at 500, 350, 250, 160, and 70~$\mu$m using the SPIRE and PACS cameras of the \textit{Herschel Space Observatory}. We used the level 2.5 data products from the Herschel Science Archive \footnote{HPPUNIMAP[B/R]: maps from the combined scan and cross-scan observation, using the \texttt{Unimap} task, suitable for analysis of extended sources; extdPxW: provide monochromatic intensities at 250, 350 and 500~$\mu$m, where PxW denotes any of the three arrays, PSW = 250~$\mu$m, PMW = 350~$\mu$m, or PLW = 500~$\mu$m} for our analysis.
%and estimate the hydrogen column density ($N(\text{H}_2)$) and dust temperature ($T_{\text{d}}$) across our targets.

The Hi-GAL data have different resolutions and plate scales at the five wavelengths of 70, 160, 250, 350, and 500~$\mu$m. Further, the data are in units of Jy~pixel$^{-1}$ at 70 and 160~$\mu$m and MJy~sr$^{-1}$ at other wavelengths. Before estimating the flux densities, we processed the data in the Herschel Interactive Processing Environment (\texttt{HIPE})\footnote{HIPE is a collaborative creation of the \textit{Herschel Science Ground Segment Consortium}, which includes ESA, the NASA \textit{Herschel Science Center}, and the HIFI, PACS, and SPIRE consortia.} and converted all data to have the same units, plate scale, and resolution. The latter was accomplished by regridding the data using a premade kernel \citep{2011PASP..123.1218A}. At the end of this process, the data at all wavelengths had a plate scale of $14\arcsec$~pixel$^{-1}$ and a resolution of $36.4\arcsec$ (which are characteristic of the 500~$\mu$m data).

To estimate the far-infrared flux density, we have used the \texttt{Photutils} package \citep{larry_bradley_2023_7946442}, an affiliated package of \texttt{Astropy}. First, the background emission is estimated and subtracted from the individual far-infrared maps by a technique known as sigma clipping, where pixels above or below a specified $\sigma$ level from the median are discarded, and the statistics are recalculated. Next, the sources are detected and extracted from the background-subtracted maps by performing image segmentation, where detected sources must have a minimum number of connected pixels that are each greater than the 3$\sigma$ value of the background noise ($\sigma$ level) in an image. Then, 
elliptical apertures are fitted on the detected sources, and the total far-infrared flux densities in each far-infrared band are estimated within the apertures.

\begin{figure}
\centering
\includegraphics[width=0.47\textwidth]{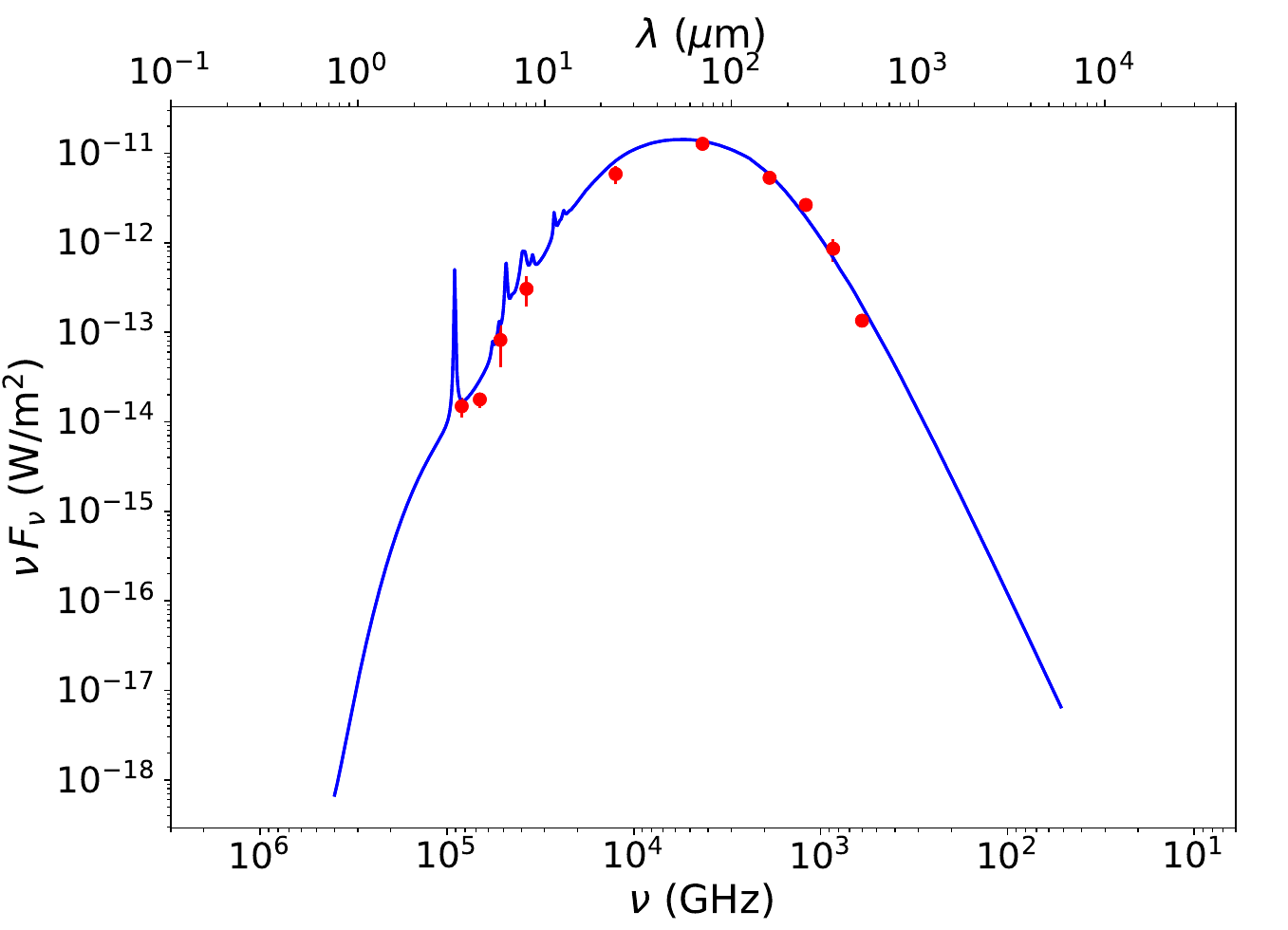}
\caption{One example SED for one of our target sources fitted using DustEM.}
\label{fig:sedsample}
\end{figure}

\begin{figure*}
\centering
\includegraphics[width=0.45\textwidth]{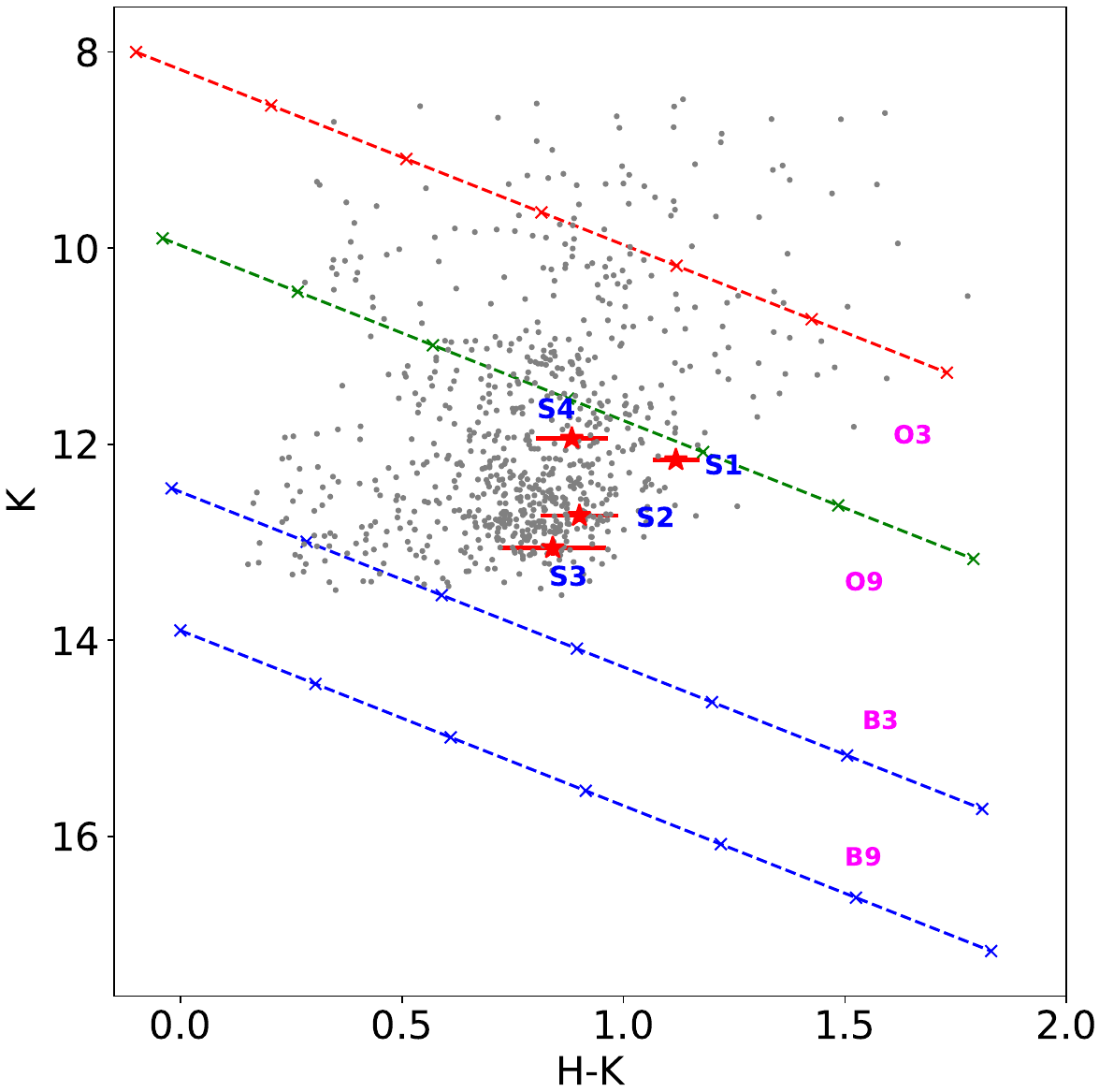}
\hfill
\includegraphics[width=0.45\textwidth]{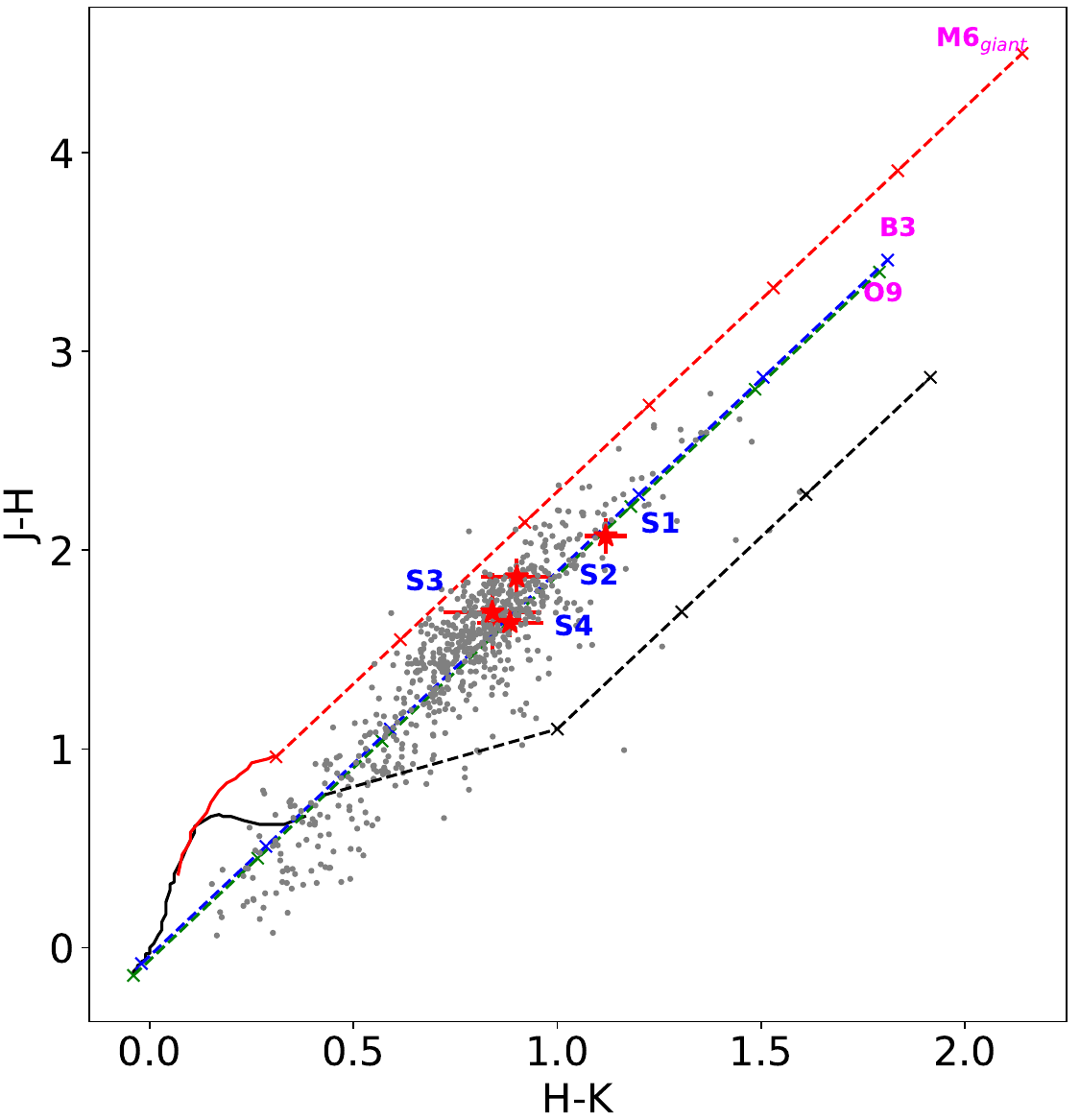}

\vspace{1cm}

\includegraphics[width=0.45\textwidth]{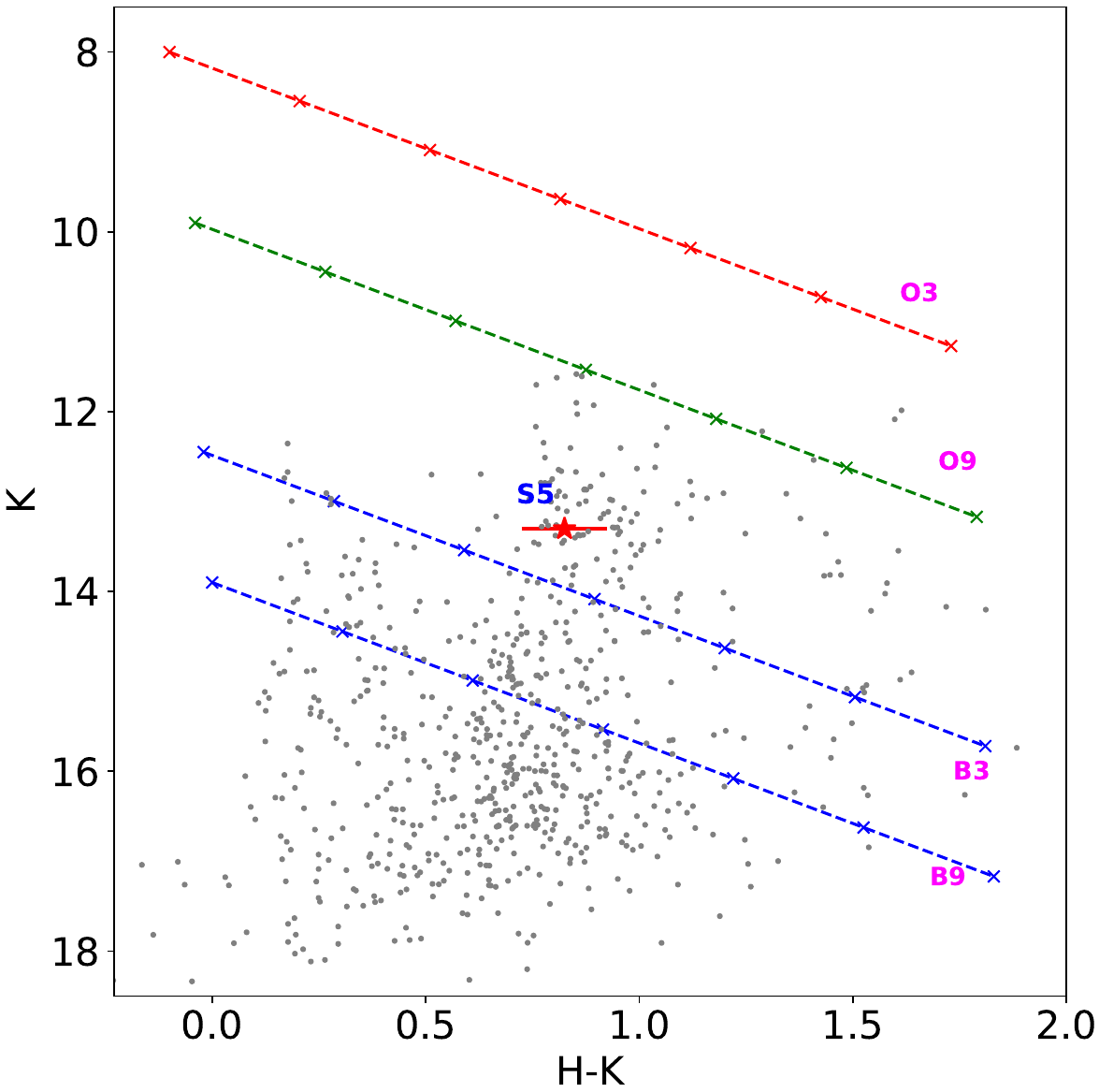}
\hfill
\includegraphics[width=0.45\textwidth]{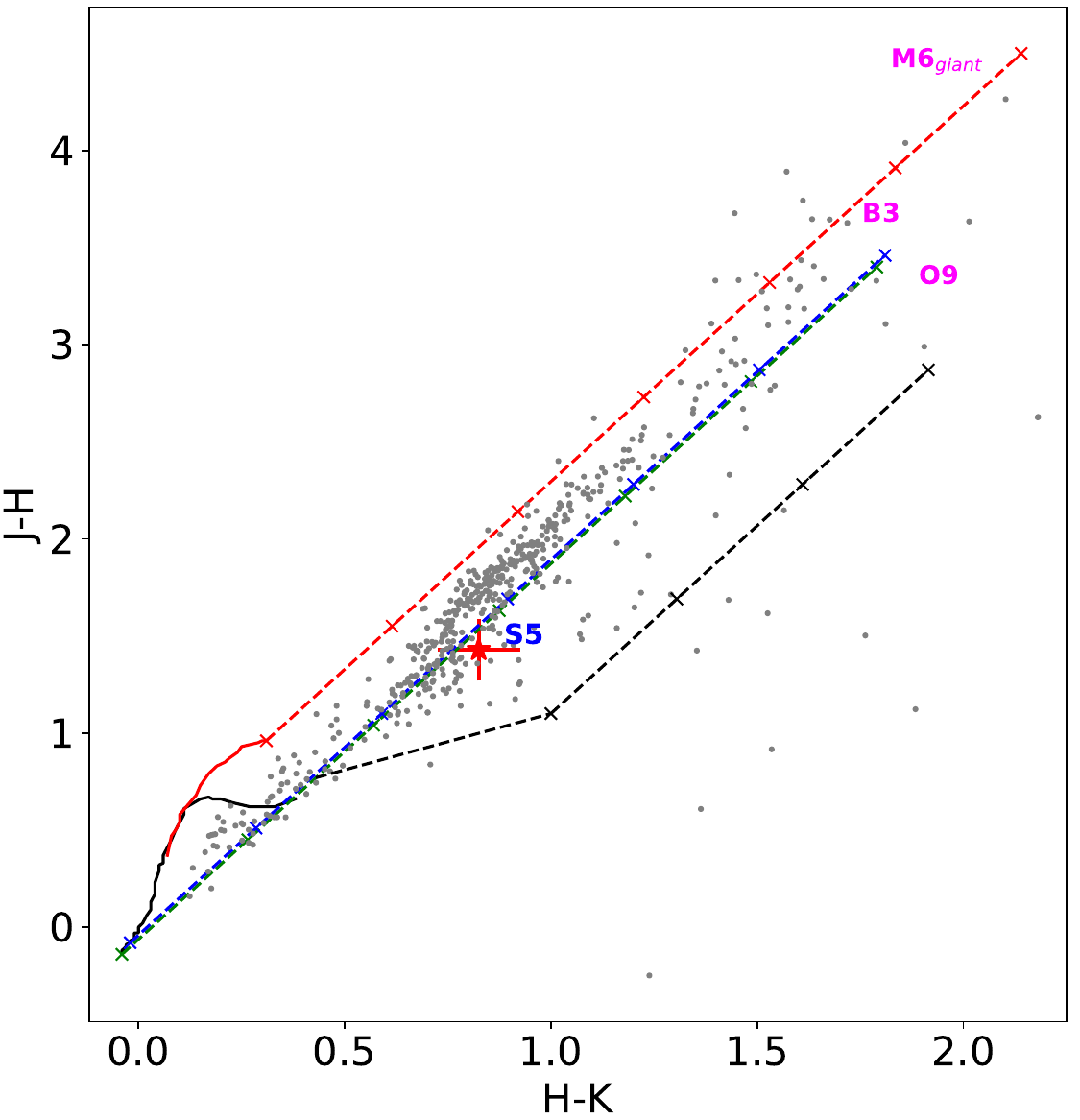}

\caption{Images showing the color-magnitude and color-color diagrams of the 2MASS and UKIDSS point sources. Top left: $K$ vs. $H-K$  color-magnitude diagram of the 2MASS near-infrared point sources within a circle of radius $4\arcmin$ centered at G19.68--0.13. The reddening vectors of the massive stars (O3--B9) are plotted on the color-magnitude diagram. Top right: $J-H$ vs. $H-K$ color-color diagram of the 2MASS point sources with spectral type earlier than B9 in the color-magnitude diagram. The solid black and red lines represent the loci of Class~V (main sequence stars) and Class~III (giant stars) objects respectively \citep{1988PASP..100.1134B, 2000asqu.book.....C}. The black long-dashed line represents the locus of the T-Tauri stars \citep{1997AJ....114..288M}. The reddening vectors corresponding to the different classes are also shown using dotted lines with crosses placed at an increasing interval of $A_{\text{V}}=5$. The candidate ionizing stars are shown using red `*' with respective error bars. Bottom row: same as the top row but for the point sources detected in the UKIDSS survey.}
\label{fig:ccd_cmd}
\end{figure*}

\subsection{Infrared spectral energy distribution}\label{analysis:dustsed}

In order to determine the Ly-photon rate from the bolometric luminosities of the H~{\small II} regions, we have complemented the Hi-GAL data with mid-infrared data from the \textit{Spitzer}-\textit{Galactic legacy infrared midplane survey extraordinaire} (GLIMPSE; \citealt{2003PASP..115..953B,2009PASP..121..213C}) and \textit{MIPS Galactic plane survey} (MIPSGAL; \citealt{2009PASP..121...76C}) surveys\footnote{\url{https://sha.ipac.caltech.edu/applications/Spitzer/SHA/}} at the \textit{Spitzer}-IRAC 3.6, 4.5, 5.8 and 8.0~$\mu$m, and \textit{Spitzer}-MIPS 24.0~$\mu$m bands. The GLIMPSE images were processed using the MOsaicker and Point source EXtractor (\texttt{MOPEX}; \citealt{makovozMopex}) package to eliminate the point sources and estimate the total flux density, including the extended emission. We have verified the accuracy of point source removal by estimating the flux density in extended emission in two ways: first, we used the method outlined above to remove the point sources and measure the total flux density after point source removal. Alternately, we performed photometry on the region covered by the H~{\small II} regions, including point sources. We then estimated the total flux density of the point sources within the aperture used for photometry using the GLIMPSE point source catalog \citep{2003PASP..115..953B}. We then subtracted this from the total flux density to estimate the flux density from extended emission. We found both methods to agree within  3--8\%, validating our approach. The mid-infrared flux densities from the GLIMPSE and MIPSGAL surveys are then combined with those from the Hi-GAL survey to construct the infrared SED of the H~{\small II} regions.

To determine the infrared luminosity of our targets, we have fit the SED through radiative transfer modeling. We have used the \texttt{DustEM} package \citep{2011A&A...525A.103C}\footnote{\url{https://www.ias.u-psud.fr/DUSTEM/}} for this purpose. The \texttt{DustEM} uses the formalism of \citet{1986A&A...160..295D} to derive the grain temperature distribution, from which the dust SED is computed for the given dust type and size distributions. The primary input parameters encompass characteristics such as the type of grains, size distribution, optical attributes, and thermal capacities of the grains, among other factors. Within this study, we have employed a dust model \citep{2011A&A...525A.103C} that consists of polycyclic aromatic hydrocarbons, amorphous carbons, and amorphous silicates. We have used the same grain size distribution and physical traits as specified in the appendix of \citealt{2011A&A...525A.103C}. A sample SED, along with the best fit from \texttt{DustEM}, are shown in Fig.~\ref{fig:sedsample}.

%However, in view of the estimation of the Ly-photon rate from the infrared fluxes, our study differs from the study of \cite{2020MNRAS.492..895D}, who have estimated the same, but by converting IRAS fluxes into a Ly-photon rate using model stellar atmospheres \citep{1973AJ.....78..929P,1986A&A...169..281C}.

\subsection{Identification of candidate ionizing stars}\label{analysis:stars}

To identify the candidate ionizing stars within the target sources, we constructed color-magnitude and color-color diagrams using photometric data of near-infrared point sources from two surveys: the \textit{Two Micron All Sky Survey} (2MASS; \citealt{2005sptz.prop20710S}) and \textit{UKIRT Infrared Galactic Plane Survey} (UKIDSS GPS; \citealt{2007MNRAS.379.1599L}). 

First, we extracted Galactic point sources from the 2MASS and the UKIDSS GPS (sixth archival data release; UKIDSSDR6plus) catalogs using a circular search radius of $4\arcmin$ centered at the coordinates given in Table~\ref{tab:source}. Next, we created color-magnitude (CMD) and color-color diagrams (CCD) using the Bessell \& Brett homogenized system \citep{2001AJ....121.2851C} for both 2MASS and UKIDSS sources. We then overplotted the loci of OB-type stars in both the CMD (assuming the distance to the UCHR as per Table~\ref{tab:source}) and CCD using the reddening law of \cite{1988PASP..100.1134B}. The Fig.~\ref{fig:ccd_cmd} shows sample CMDs and CCDs along with reddening vectors with crosses placed at intervals of $A_{\text{V}}=5$. The stars that are located in both the CMD and CCD with colors expected for OB-type stars and with the same visual extinction in both diagrams are selected as candidate ionizing stars.

\begin{figure}
\centering
\includegraphics[width=0.45\textwidth]{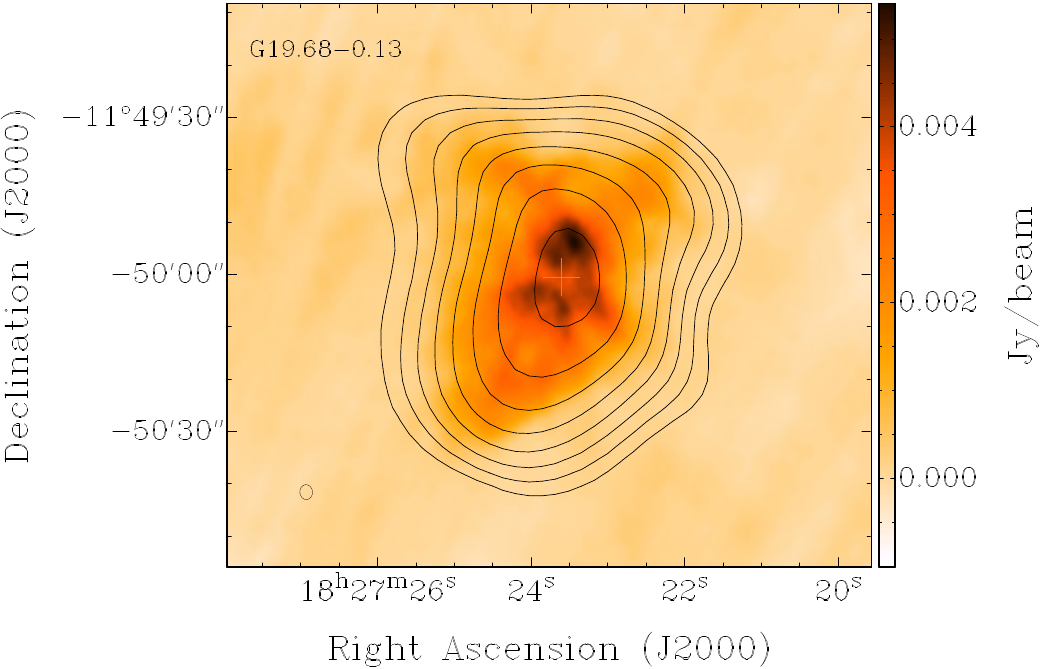}
\caption{uGMRT radio continuum map of G19.68$-$0.13 overlaid with the radio contours from the GLOSTAR-D continuum map in black. The contours have started at the 3$\sigma$-level flux and have increased in steps of $\sqrt{2}$. The location of G19.68$-$0.13 reported in the THOR radio continuum catalog is shown using a green `$+$' sign. The corresponding beam size is shown at the bottom-left corner of the figure.}
\label{fig:radioemission}
\end{figure}

\begin{figure*}
\centering
\includegraphics[width=0.50\textwidth]{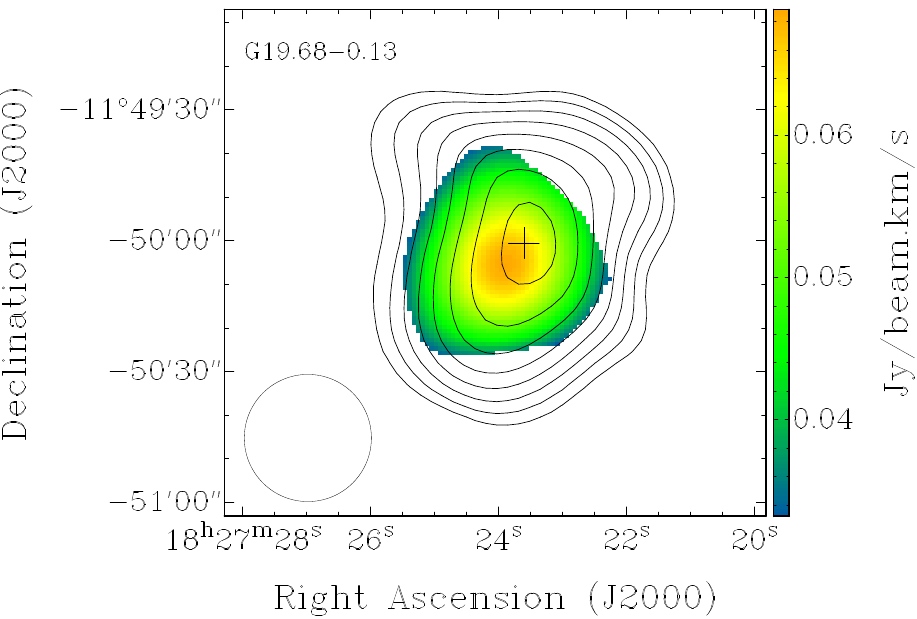}
\hfill
\includegraphics[width=0.43\textwidth]{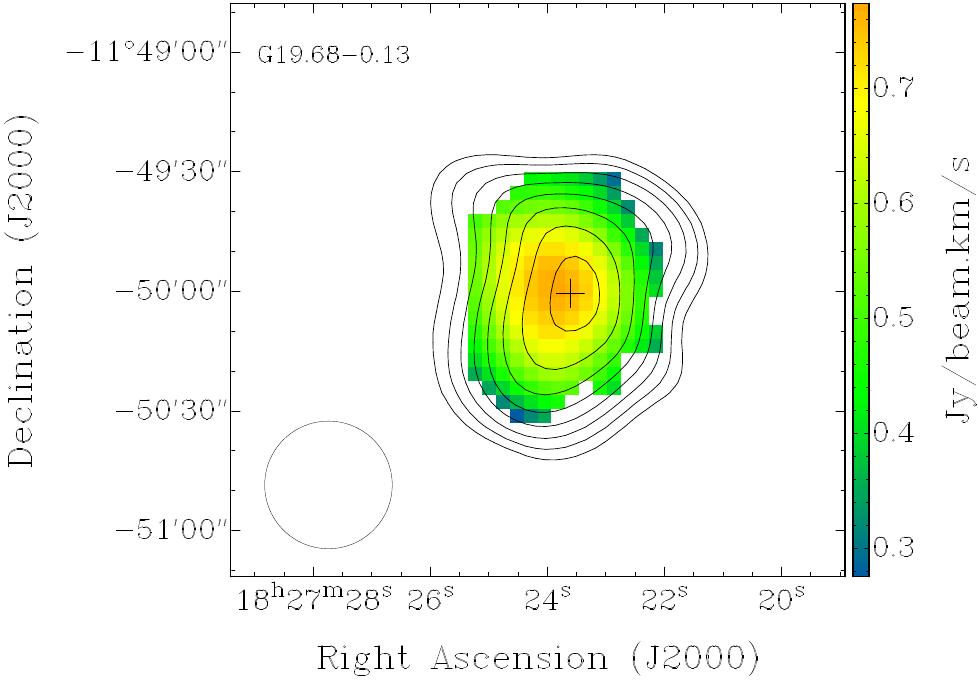}

\caption{Images showing the moment-0 maps of G19.68$-$0.13 overlaid with the radio contours from the GLOSTAR-D continuum maps in black. Left: uGMRT, right: GLOSTAR-D. The contour levels are similar to those in Fig.~\ref{fig:radioemission}. The location of G19.68$-$0.13 reported in the THOR radio continuum catalog is shown using a black `$+$' sign. The respective beam sizes are shown at the bottom-left corners of the figures.}
\label{fig:radiom0}
\end{figure*}

%--------------------------------------------------------------------

\section{Results}\label{sec:results}

\subsection{Radio emission}\label{sec:radioresults}

\subsubsection{Radio continuum}\label{resradcont}

We detected extended emission around all the sources in both uGMRT and GLOSTAR (GLOSTAR-D and GLOSTAR-D+Eff) observations. Figure~\ref{fig:radioemission} shows an example of radio continuum emission at 2$\arcsec$ resolution using the uGMRT with the 18$\arcsec$ resolution GLOSTAR-D emission overlaid in contours (the images for the full sample are provided in the left panel of Fig.~\ref{fig:radioemission-app}). Notably, there are no morphological differences between the continuum images at 1.35 GHz and 5.79 GHz.

The integrated flux densities of the sources in the GLOSTAR-D+Eff maps are typically larger than that seen in GLOSTAR-D alone (Table~\ref{tab:radioresults}), indicating that some emission is resolved in the latter. %However, in one or two sources, the larger flux density is an outcome of contamination from neighboring sources, and these are discussed in Sect.~\ref{sec:notes}.

%\subsection{Radio emission}\label{sec:radioresults}

%The targets in our sample are studied in radio wavelengths using radio continuum maps constructed from the uGMRT, GLOSTAR-D, GLOSTAR-B, and GLOSTAR-D+Eff data. Notably, there are no morphological differences between the 1.35~GHz and 5.79~GHz continuum images from uGMRT and GLOSTAR-D (Fig.~\ref{fig:radioemission} and left panels of Fig.~\ref{fig:radioemission-app}). The integrated flux densities of the H~{\small II} regions within the 3$\sigma$ level in uGMRT (1.35~GHz), GLOSTAR-D (5.79~GHz), and GLOSTAR-D+Eff (5.85~GHz) are tabulated in Table~\ref{tab:radioresults}.

%The GLOSTAR-D+Eff radio continuum maps show that the integrated flux densities in these maps are typically larger (by almost a factor of 3 in one source) than that in the GLOSTAR-D maps (Table~\ref{tab:radioresults}), as the former will include the full extent of extended emission since they incorporate zero-spacing information. However, in one or two sources, this includes emissions from neighboring sources with a shell-like morphology. These are discussed in Sect.~\ref{sec:notes}.

\subsubsection{Spectral index}\label{resradspec}

Table~\ref{tab:radioresults} shows the in-band spectral indices from the GLOSTAR-D data around the location of the peak emission and the inter-band spectral indices estimated from the integrated flux densities of the sources in uGMRT and GLOSTAR-D. We find that the spectral index ($\alpha$; $S_\nu \propto \nu^{\alpha}$) is close to $-0.1$ in all cases, indicating that the emission is thermal and optically thin at frequencies greater than 1.35 GHz.

However, the radio continuum images at high resolution show the emission to be clumpy, with several sources showing pockets of compact emission that are surrounded by diffuse extended emission. To explore the nature of compact emission, we have examined the spectral index using GLOSTAR-B data, which is only sensitive to emission on scales smaller than 4$\arcsec$. We detect pockets of compact emission at the 3$\sigma$-level or greater in the GLOSTAR-B maps for four out of eight sources (G22.76$-$0.48, G24.47$+$0.49, G25.69$+$0.03, and G28.30$-$0.39). The GLOSTAR-B in-band spectral index was found to range from 1.21 to 1.97 for these pockets; however, the uncertainty in the spectral index is high (often exceeding 100\%) due to the low signal-to-noise ratio. Although the high uncertainties are a caveat, a positive spectral index up to the frequency of GLOSTAR surveys requires emission measures greater than $\sim 10^8$~pc~cm$^{-6}$, which coupled with their compact sizes ($\lesssim$ 0.1 pc) is indicative of electron densities that are found in ultracompact H~{\small II} regions \citep{2005IAUS..227..111K}.

We have also determined the inter-band spectral index for pockets of compact emission that are detected in both the uGMRT observations and the \textit{Co-Ordinated Radio `N' Infrared Survey for High-mass star formation} (CORNISH; \citealt{2012PASP..124..939H}) survey. We find the spectral index to be positive (even after accounting for uncertainties) in all cases, showing that the compact emission is optically thick thermal emission. %\textbf{However, except for the pockets having GLOSTAR-B counterparts, others have sizes much larger than 0.1 pc, the typical size criteria for an H~{\small II} region to be a UCHR \citep{2005IAUS..227..111K}.}

The results above are characteristic of hierarchical density structure in H~{\small II} regions, similar to that observed by \citet{2002ASPC..267..373K} and \citet{2021A&A...645A.110Y}. Although many H~{\small II} regions are observed to have pockets of compact emission whose electron densities are similar to that found in UCHRs, the overall emission is dominated by the diffuse emission that is optically thin down to 1.35 GHz or lower. Thus, high-resolution observations that are not sensitive to extended emission (e.g., WC89, CORNISH, GLOSTAR-B) are dominated by the compact emission components that have a positive spectral index, whereas the spectral index from lower resolution observations (e.g., GLOSTAR-D, integrated emission from uGMRT) is close to $-0.1$.

\subsubsection{Source classification}\label{ressoucla}

As indicated in Sect.~\ref{sec:intro}, H~{\small II} regions are classified based on their size, electron density, spectral index, and emission measure. However, the hierarchical density structure makes the classification of the H~{\small II} regions in our sample challenging. %\textbf{So, as discussed in the previous section, we searched for the dense compact components within our target sources using the GLOSTAR-B data}.
Based on the detected compact emission alone (at the 3$\sigma$-level of GLOSTAR-B data and the CORNISH survey), four out of eight sources (G22.76$-$0.48, G24.47$+$0.49, G25.69$+$0.03, and G28.30$-$0.39) would be classified as UCHRs, according to their sizes ($\lesssim$ 0.1~pc) and emission measures ($\gtrsim$ $10^7$~pc~cm$^{-6}$), with G24.47$+$0.49 and G25.69$+$0.03 being already classified as UCHRs in the catalogs of \citet{2013ApJS..205....1P} and \citet{2020MNRAS.492..895D} (see Sect.~\ref{sec:notes} for more details).
%\textbf{We have also detected their CORNISH counterparts at the 3$\sigma$-level ($\sigma$ = 0.3--0.4 mJy~beam$^{-1}$). Hence, we attempted to estimate a better spectral index using the 3$\sigma$-level flux densities from the CORNISH survey and that measured from the combined image of GLOSTAR-B sub-bands. Although the resulting spectral indices still have large uncertainties, we find a lower limit that is still positive. Hence, we treat these four sources as UCHRs in our work. The left panel of Fig.~\ref{fig:radioemission-app} shows the locations of the detected UC cores (in this work and also in the literature) surrounded by extended emission (see the caption for details).}

The compact emission from the other four sources (G19.68$-$0.13, G20.99$+$0.09, G27.49$+$0.19, and G28.81$+$0.17) is seen to be fragmented in the CORNISH maps, which is an artifact of the \texttt{CLEAN} process (during imaging of the interferometric data) when the resolution is significantly higher than the scale of extended emission. The overall size of the envelope that encompasses all the fragmented emission is found to range from 0.2 to 1.2 pc. Thus, following the `size' criteria, these H~{\small II} regions cannot be classified as UCHRs. Hence, a classification scheme based on the size and spectral index of dense emission alone would result in the sample comprising four UCHRs and four compact H~{\small II} regions. However, if the extended emission is included, all eight sources would be classified as compact H~{\small II} regions. It is to be noted that many early surveys that classified Galactic H~{\small II} regions as UCHRs have been carried out at high angular resolution with relatively poor sensitivity to extended emission (e.g. WC89, CORNISH). The semantics of how to classify H~{\small II} regions that have a hierarchical structure with extended emission that co-exists with compact emission is an open question. Although the presence of extended emission presents difficulties in the classification of H~{\small II} regions, it plays an important role in resolving the outstanding problems associated with UCHRs, as demonstrated in Sect.~\ref{sec:discuss}.

\subsubsection{RRL emission}\label{resradlin}

In addition to the radio continuum maps, we have RRL data cubes of the target regions at an angular resolution of 25$\arcsec$. These cubes are used to trace the dynamics of ionized gas, effects of feedback from H~{\small II} regions on the surrounding environment, the electron temperature (e.g., \citealt{2006ApJ...653.1226Q}), etc. In addition, the line velocity can be used to estimate the kinematic distance to the source.

\begin{figure}
\centering
\includegraphics[width=0.45\textwidth]{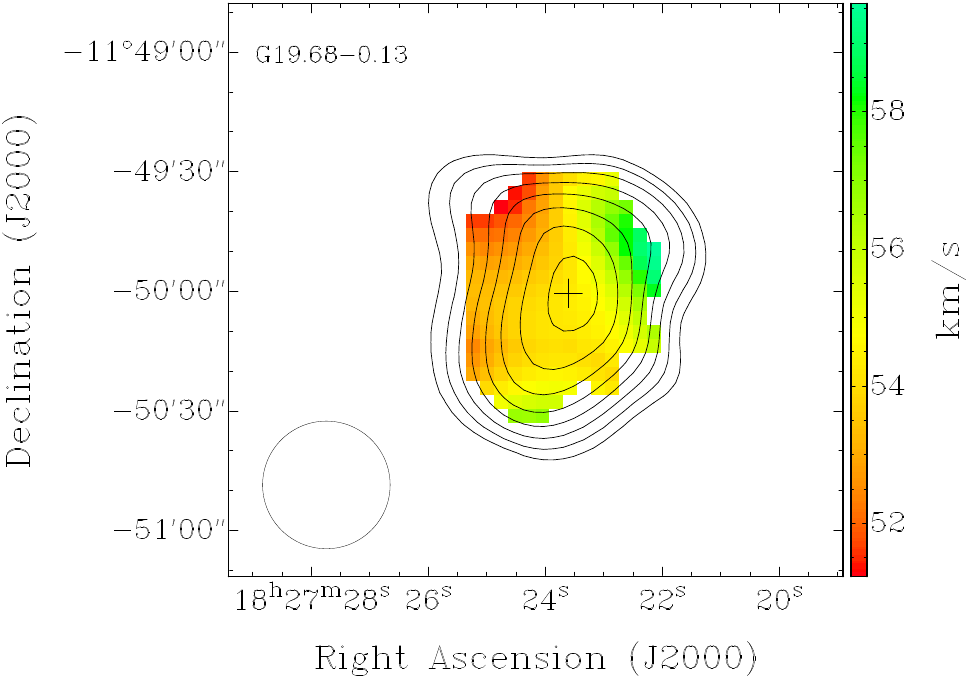}

\caption{RRL peak velocity distribution map of G19.68$-$0.13 from the GLOSTAR-D data overlaid with the GLOSTAR-D radio continuum contours in black. The contour levels are similar to those in Fig.~\ref{fig:radioemission}. The location of G19.68$-$0.13 reported in the THOR radio continuum catalog is shown using black `$+$', and the corresponding beam size is shown at the bottom-left corner of the figure.}
\label{fig:velocityfield}
\end{figure}

The left and right panels of Fig.~\ref{fig:radiom0} show the moment-0 or velocity-integrated maps of G19.68$-$0.18 from the uGMRT and GLOSTAR-D, respectively. Although the two datasets are expected to trace different gas densities, the distribution of RRL emission is seen to be very similar in both maps. Figure~\ref{fig:velocityfield} shows the velocity field of the ionized gas using GLOSTAR-D, obtained by fitting a Gaussian profile to the RRL line at each pixel that has a signal-to-noise ratio greater than 3. The velocity field using the uGMRT data is similar but at a lower signal-to-noise ratio due to the lower line intensities at higher Hn$\alpha$ transitions. Although limited by the signal-to-noise ratio, the lack of any significant differences in either the extent of RRL emission or the velocity field between uGMRT and GLOSTAR-D suggests that both datasets trace the same ionized gas. The moment-0 maps and velocity field for the full sample are shown in Fig.~\ref{fig:radioemission-app} (right panel), and Fig.~\ref{fig:velocityfield-app}, respectively, and the velocity fields of individual sources are discussed in Sect.~\ref{sec:notes}. The line widths of the RRLs range from 24 to 30~km~s$^{-1}$, which is typical of compact and ultracompact H~{\small II} regions (e.g., WC89, \citealt{2001ApJ...549..979K}).

%In addition to the radio continuum maps, we have RRL data cubes of the target regions. These cubes are used to trace the dynamics of ionized gas, effects of feedback from H~{\small II} regions on the surrounding environment, the electron temperature (e.g., \citealt{2006ApJ...653.1226Q}), etc. In addition, the line velocity can be used to estimate the kinematic distance to the source. The moment-0 or velocity-integrated intensity maps of G19.68$-$0.13 generated from the GLOSTAR-D and uGMRT RRL data cubes are shown in the right panel of Fig.~\ref{fig:radiom0}, whereas Fig.~\ref{fig:velocityfield} shows the velocity field of the ionized gas across G19.68$-$0.13 using the GLOSTAR-D data (Fig.~\ref{fig:radioemission-app} and Fig.~\ref{fig:velocityfield-app} contain the rest). The moment-0 maps trace a large extent of diffuse gas due to the reasons discussed earlier in Sect.~\ref{sec:intro}. The velocity field obtained using the uGMRT data is similar but at a lower signal-to-noise ratio. These maps are generated by fitting a Gaussian profile to each pixel of the spectral cube that has a signal-to-noise ratio greater than 3. The line widths obtained from our fits range from 24 to 30 km~s$^{-1}$, which is typical of UCHRs (e.g. \citealt{1989ApJS...69..831W, 2001ApJ...549..979K}). The velocity fields of individual sources are described in Sect.~\ref{sec:notes}. 

\subsection{Electron temperature}\label{eltemp}

We have used the RRL line-to-radio continuum ratio to estimate the electron temperature ($T_{\text{e}}$) in the target H~{\small II} regions (see Appendix~\ref{appendix} for details on the methodology). The $T_{\text{e}}$ in our sample ranges from 7200~K to 9600~K (see Table~\ref{tab:radioresults}). A visual inspection of the data shows an increasing $T_{\text{e}}$ with an increase in the galactocentric distance of our sample. Our results are broadly consistent (within 300--400~K) with the relation between $T_{\text{e}}$ and galactocentric radius (e.g., \citealt{2006ApJ...653.1226Q}) that is a consequence of the metallicity of the Galactic disk decreasing with galactocentric radius. However, for G19.68$-$0.13 and G27.49$+$0.19, our estimated $T_{\text{e}}$ values are within 900--1000~K of the relation given in \citet{2006ApJ...653.1226Q}.

\begin{sidewaystable*}
\caption{Properties of the radio continuum emission using uGMRT and GLOSTAR data.}
\label{tab:radioresults}
\centering
\begin{tabular}{c c c c c c c c c c c c}
\hline\hline 
Source name & S$_\nu$ (1.35~GHz) & S$_\nu$ (5.79~GHz) & S$_\nu$ (5.85~GHz) & $\log(N_{\text{Ly}})$ & $\log(N_{\text{Ly}})$ & ZAMS & $T_{\text{e}}$ & $D_{\text{UC}}$ & $\alpha_{\text{UC}}$ & $\alpha_{\text{p}}$ & $\alpha$ \\
 & (Jy) & (Jy) & (Jy) & (1.35~GHz) & (5.85~GHz) &  & (K) & (pc) &  &  & \\
\hline
$^{*}$G19.68$-$0.13 & $0.622 \pm 0.012$ & $0.481 \pm 0.018$ & 0.57 $\pm$ 0.07 & 48.82 & 48.85 & O6.5 & 9627 & -- & -- & --0.070 & --0.098 \\
$^{*}$G20.99$+$0.09 & $1.344 \pm 0.014$ & $0.630 \pm 0.017$ & 1.17 $\pm$ 0.10& 47.54 & 47.55 & B0 & 9162 & -- & -- & --0.132 & --0.156 \\
$\;\,$G22.76$-$0.48 & $1.788 \pm 0.025$ & $0.672 \pm 0.013$ & 1.81 $\pm$ 0.11 & 48.56 & 48.63 & O7 & 8216 & 0.08 & $1.97 \pm 2.19$ & --0.136 &  --0.144 \\
$\;\,$G24.47$+$0.49 & $3.178 \pm 0.006$ & $3.502 \pm 0.012$ & 3.76 $\pm$ 0.15 & 48.98 & 49.12 & O6 & 7956 & 0.11 & $1.96 \pm 1.98$ & --0.121 & --0.129 \\
$\;\,$G25.69$+$0.03 & $0.883 \pm 0.016$ & $0.533 \pm 0.019$ & 0.83 $\pm$ 0.06 & 49.02 & 49.06 & O6 & 7680 & 0.04 & $1.66 \pm 0.38$ & --0.126 & --0.137 \\
$^{*}$G27.49$+$0.19 & $1.701 \pm 0.014$ & $0.785 \pm 0.011$ & 1.47 $\pm$ 0.08 & 47.87 & 47.87 & O9.5 & 7213 & -- & -- & --0.139 & -0.151 \\
$\;\,$G28.30$-$0.39 & $0.820 \pm 0.015$ & $0.654 \pm 0.010$ & 0.71 $\pm$ 0.07 & 48.80 & 48.80 & O6.5 & 8829 & 0.11 & $1.21 \pm 5.60$ & --0.113 & -0.132 \\
$^{*}$G28.81$+$0.17 & $1.462 \pm 0.012$ & $0.971 \pm 0.014$ & 1.30 $\pm$ 0.09 & 48.89 & 48.90 & O6.5 & 8276 & -- & -- & --0.125 & -0.147 \\
\hline
\end{tabular}
\tablefoot{\tablefoottext{*}{sources comprise of multiple fragments of compact emission,} $D_{\text{UC}}$ = size of the UCHR (excluding the extended emission) using the GLOSTAR-B data, $\alpha_{\text{UC}}$ = spectral index of the UCHR (excluding the extended emission) using the GLOSTAR-B data, $\alpha_{\text{p}}$ = spectral index using the in-band GLOSTAR-D flux densities within a grid of $3\times3$ pixels centered at the pixel with the peak emission, $\alpha$ = spectral index estimated using the uGMRT and GLOSTAR-D integrated flux densities.}
\end{sidewaystable*}

\begin{figure}
\centering
\includegraphics[width=0.45\textwidth]{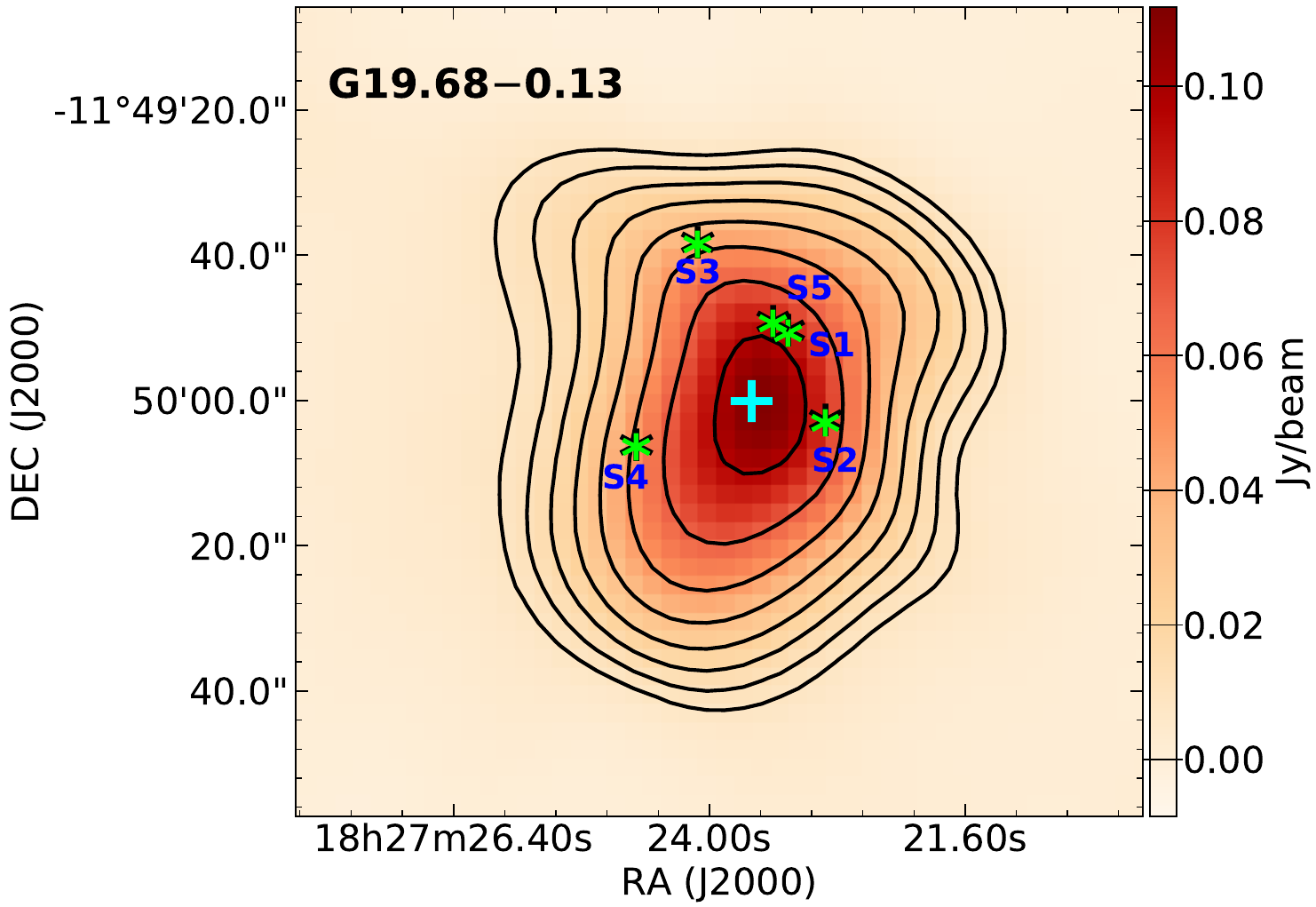}
\caption{Locations of the candidate ionizing stars (labeled using an S\_\_ format) are shown using green `*' signs for G19.68$-$0.13 (see Table~\ref{tab:stars} for details). The cyan `+' indicates the location of G19.68$-$0.13 reported in the THOR radio continuum catalog.
}\label{fig:starlocations}
\end{figure}

\subsection{Energetics of the target H~{\small II} regions}\label{sec:energy}

The Ly-photon rate can be inferred from the measured radio flux density using the following equation \citep{2016A&A...588A.143S}
\begin{equation}
    N_{\text{Ly}} \geq 4.76 \times 10^{42} \; \nu^{0.1} \; d^2 \; S_\nu \; {T_{\text{e}}}^{-0.45},
    \label{eq:eq1}
\end{equation}
where $S_{\nu}$ is the flux density in Jy, $\nu$ is the frequency in GHz, $d$ is the distance to the source in pc, and $T_{\text{e}}$ is the electron temperature of the ionized gas in K. The Ly-photon rates computed from uGMRT and GLOSTAR-D+Eff are shown in Table~\ref{tab:radioresults}. There is good agreement between the Ly-photon rates computed from the two surveys, with the rate inferred from GLOSTAR-D+Eff being slightly larger than that from uGMRT in a few sources. This confirms that both observations mostly trace the same optically thin diffuse emission. It also shows the efficiency of uGMRT in picking extended emission due to its central array comprising 14 antennas within a 1 sq km region.

Assuming that all the ionizing radiation comes from a single main sequence star, one can estimate the spectral type of the ionizing star. We have used the Ly-photon rate from the GLOSTAR-D+Eff observations and the formalism in \citet{2005A&A...436.1049M} for this estimation (7th column of Table~\ref{tab:radioresults}). We have also determined the candidate ionizing stars in each H~{\small II} using the NIR CCD and CMD (see Sect.~\ref{analysis:stars}), the locations of which are shown in Fig.~\ref{fig:starlocations}. We see that the ionizing radiation comes from multiple stars for all H~{\small II} regions in our sample, and we find reasonable agreement between the Ly-photon rate estimated from radio emission and the total rate from all candidate ionizing stars for our target sources (see Table~\ref{tab:stars}). This important observation, along with the detection of diffuse radio emission, indicates that both the dense and extended emission have originated from the same group of candidate ionizing sources. Although it is well established that massive stars form in clusters, the presence of multiple candidate ionizing stars in our targets reaffirms this phenomenon.

\begin{table*}
\caption{Details of the candidate ionizing stars.}
\label{tab:stars}
\centering
\begin{tabular}{c c c c c c}
\hline\hline
Source name & Spectral type$^{\text{a}}$  & No. of Stars$^{\text{b}}$ & Total $\log(N_{\text{Ly}})^{\text{c}}$ & Total $\log(N_{\text{Ly}})^{\text{d}}$ & $\log(N_{\text{Ly}})$  \\
 &  &  & (lower limit) & (upper limit) & (D+Eff) \\
\hline
G19.68$-$0.13 & $>$ O9 & 1 & 48.08 & 48.85 & 48.85     \\
  & $>$ B3 but $<$ O9 & 4 &  &  &      \\
\hline
G20.99$+$0.09 & $>$ B3 but $<$ O9 & 5 & 44.53 & 48.78 & 47.55     \\
  & $\sim$ B3 & 2 &  &  &      \\
\hline
G22.76$-$0.48 & $>$ O9 & 1 & 48.08 & 49.96 & 48.63     \\
  & $>$ B3 but $<$ O9 & 6 &  &  &      \\
  & $\sim$ B3 & 1 &  &  &      \\
\hline
G24.47$+$0.49 & $\sim$ O9 & 2 & 44.89 & 48.98 & 49.12    \\
  & $>$ B3 but $<$ O9 & 6 &  &  &      \\
  & $\sim$ B3 & 10 &  &  &      \\
\hline
G25.69$+$0.03 & $\sim$ O9 & 1 & 44.17 & 48.38 & 49.06     \\
  & $>$ B3 but $<$ O9 & 1 &  &  &      \\
  & $\sim$ B3 & 2 &  &  &      \\
\hline
G27.49$+$0.19 & $>$ B3 but $<$ O9 & 2 & 43.99 & 48.38 & 47.87     \\
\hline
G28.30$-$0.39 & $>$ B3 but $<$ O9 & 2 & 44.17 & 48.38 & 48.80     \\
  & $\sim$ B3 & 1 &  &  &      \\
\hline
G28.81$+$0.17 & $>$ B3 but $<$ O9 & 4 & 44.38 & 48.68 & 48.90     \\
  & $\sim$ B3 & 1 &  &  &     \\
\hline
\end{tabular}
\tablefoot{\tablefoottext{a}{spectral type range of the candidate ionizing stars including uncertainty in magnitudes and colors,} \tablefoottext{b}{total number of candidate ionizing stars in each spectral type category,} \tablefoottext{c}{total Ly-photon rate calculated from all candidate stars considering them to be the latest spectral type in the category,} \tablefoottext{d}{same as c but considering them to be the earliest spectral type in the category.}}
\end{table*}

\subsection{Infrared luminosity}

As explained in Sect.~\ref{analysis:dustsed}, we have used data from the GLIMPSE, MIPSGAL, and Hi-GAL surveys to construct the SED of dust emission and modeled it using the DustEM package. The infrared luminosity ($L^{\text{\,dust}}_{\text{IR}}$) of the H~{\small II} regions ranges from $6.92 \times 10^3$ to $1.95 \times 10^5~L_\odot$. This is consistent with the study by \citet{2015A&A...579A..71C}, who estimated the infrared luminosity (21~$\mu$m to 1.1~mm) of $\sim$ 200 young H~{\small II} regions to range from $10^3$ to $10^6~L_\odot$. Similar to Sect.~\ref{sec:energy}, we have determined the total luminosity from the ionizing stars and compared it with the luminosity determined by fitting the SED. We find the values to be consistent, reaffirming that the ionization in the H~{\small II} regions is from a cluster of massive stars rather than a single star.

%--------------------------------------------------------------------

\section{Notes on individual sources}\label{sec:notes}

\textit{G19.68$-$0.13} -- We have detected arcmin scale extended emission surrounding the compact emission, which appears to be fragmented in the CORNISH radio continuum map at 5~GHz. These fragments altogether span an area of size $\approx$ 0.9~pc. A small velocity gradient (52--58~km~s$^{-1}$) is also present from the east to west direction. Five candidate ionizing stars are also detected, with three stars, including one O9-type star, close to the reported location in Table~\ref{tab:source}.

\textit{G20.99$+$0.09} -- This source exhibits diffuse emission that extends out to $1\arcmin-1.25\arcmin$ from the center. The high-resolution map from uGMRT shows at least three compact emissions embedded within the extended emission. Moreover, the CORNISH 5~GHz map also shows similar morphology with their combined size $\approx$ 0.8~pc. The velocity field shows a gradient from the south-west to north-east direction. The data from the \textit{$^{12}$CO (3--2) High-Resolution Survey} (COHRS; \citealt{2013ApJS..209....8D}) of the molecular gas also shows similar velocity distribution, suggesting that the velocity field seen in the ionized gas is derived from that of the parent molecular cloud. We also detect several candidate ionizing stars within this H~{\small II} region, indicating prior star formation activity within the parent cloud.

\textit{G22.76$-$0.48} -- This source has a complex morphology with the extended emission reaching out to $\sim 4\arcmin$ in the GLOSTAR-D+Eff map. However, the extended emission is well within the boundary of far-infrared emission arising from the cold dust. The GLOSTAR-B radio continuum map shows a 7$\sigma$-level source near the location reported in the THOR radio continuum catalog. We have also detected a corresponding CORNISH source at this location. The spectral index is found to be positive (translates to an emission measure $\gtrsim$ $10^7$~pc~cm$^{-6}$) along with the size $\leq$ 0.1~pc. These observations are consistent with this source being identified as a UC core. Moreover, the candidate ionizing stars are spread across multiple regions in the radio continuum maps, suggesting that high-mass star formation is happening in multiple groups within this H~{\small II} region. An O4--O7 type star is responsible for the bulk of ionizing radiation \citep{2014A&A...569A..20M}, with smaller contributions from multiple other B-type stars. We also detect a couple of young stellar objects (YSOs) near the south-west region of the radio continuum map, where no candidate ionizing stars are detected. The velocity field in this source is complex, with lower velocities being seen in the central region. Also, there is a small extension of ionized gas from the east towards the south of this H~{\small II} region. However, We couldn’t find any \textit{APEX Telescope Large Area Survey of the Galaxy} (ATLASGAL; \citealt{2009A&A...504..415S,2018MNRAS.473.1059U}) clump candidates associated with this emission, whereas multiple ATLASGAL clumps (located at a similar distance) are associated with other parts of this H~{\small II} region. We were also unable to find any candidate ionizing stars or YSOs in this region. This, coupled with the non-detection of any RRL emission, leads us to not consider this small extension as being associated with this source.

\textit{G24.47$+$0.49} -- This H~{\small II} region displays a fragmented shell morphology with significant extended emission detected in both uGMRT and GLOSTAR, with the UC core detected (listed in the UCHR catalog of \citealt{2013ApJS..205....1P}) near the reported position in the THOR catalog. The GLOSTAR-B map also shows a 3$\sigma$-level source at this location matching with the criteria of a UC core. Moreover, the GLOSTAR-D and GLOSTAR-D+Eff maps show multiple bubbles surrounding the central shell-like structure, with a similar structure seen in the GLIMPSE survey at 8.0~$\mu$m. \citet{2019MNRAS.488.1141J} have identified three bubbles by fitting ellipses (major axis $\approx$ $1\arcmin$) to those structures, and the fragmented shell appears to have formed at the intersection point of these bubbles. However, since RRLs are not detected in these bubbles, we cannot determine whether they are directly associated with the UCHR or whether they trace an earlier generation of star formation. We have hence assumed the extent of this source to be equal to that detected using uGMRT (i.e. excluding the bubbles). The flux density measurements in the GLOSTAR-D and GLOSTAR-D+Eff maps are performed within this assumed area. We have identified two candidate O-type stars and a supergiant star (S39) using data from the 2MASS, UKIDSS, and \textit{Panoramic Survey Telescope and Rapid Response System} (Pan-STARRS; \citealt{2016arXiv161205560C}) surveys. One of the candidate O-type stars is identified in one of the fragments, and a candidate B-type star is located near the center of the shell.

\textit{G25.69$+$0.03} -- The extended emission shows a cometary morphology (especially in the high-resolution uGMRT map) with high-velocity ionized gas near the head of the comet. However, the UC core (identified as a 3$\sigma$-level source with size $\leq$ 0.1~pc from the GLOSTAR-B map along with a 3$\sigma$-level CORNISH counterpart) is located at the edge of the extended emission. Two candidate ionizing stars (one of which has spectral type O9) along with multiple other stars of late spectral types are also detected near the head using the 2MASS, \textit{Naval Observatory Merged Astrometric Dataset} (NOMAD; \citealt{2004AAS...205.4815Z}), and \textit{The Guide Star Catalog} \citep{2008AJ....136..735L} data. However, the Ly-photon rate estimated from the GLOSTAR-D+Eff radio continuum emission is higher than the total contribution from all detected candidate ionizing stars. This may suggest that a few more such stars are not detected in those surveys. The location of the ionizing stars significantly away from the UC core suggests that the UC core is not the source of extended emission for this region. The UC core may be a case of triggered star formation, although it requires further attention in future studies. Two more H~{\small II} regions are also located towards the north. However, they are not related to our target.

\textit{G27.49$+$0.19} -- From the radio continuum morphology, this H~{\small II} region appears to be bipolar with two ionized bubbles towards the north and south direction. The $^{12}$CO (3--2) molecular emission data from the COHRS survey shows emission surrounding these ionized bubbles. However, \cite{2018ApJ...867..167L} have not detected any molecular outflow using the COHRS survey data. The compact emission is fragmented in the high-resolution map of CORNISH over a size of $\approx$ 0.2~pc. The signal-to-noise of the RRL map allows us to construct the velocity field only in the central region, with no apparent dichotomy between the northern and southern parts. Two candidate ionizing stars are seen near this central region, located at the meeting point of those two ionized bubbles.

\textit{G28.30$-$0.39} -- This source shows a slight offset between the peak of the moment-0 map and that of the radio continuum. There are two peaks in the far-infrared maps from Hi-GAL containing two H~{\small II} regions. However, the H~{\small II} region within the westwards peak is located at a different distance \citep{2018MNRAS.473.1059U}. Hence, the corresponding radio contours are not shown here. Three candidate ionizing stars, along with a YSO (the YSO and a candidate ionizing star are located near the compact emission), are also detected in this H~{\small II} region, although none of them are at the peak of the radio emission. The location of the UC core is identified from the GLOSTAR-B and CORNISH maps (3$\sigma$ level), and it is located 18$\arcsec$ away from the coordinates reported in the THOR radio continuum catalog. Also, this source is very weak in both the high-resolution maps. Moreover, the Ly-photon rate inferred from the GLOSTAR-D+Eff radio continuum emission is greater than the cumulative contribution from all candidate stars detected using near-infrared wavelengths (see Table~\ref{tab:stars}). This could be due to the non-detection of the entire population of candidate ionizing stars within this H~{\small II} region.

\textit{G28.81$+$0.17} -- At radio wavelengths, the compact emission appears to be fragmented in the CORNISH map, spanning across an area of $\approx$ 1.2~pc. There are two candidate ionizing stars near the peak of the radio emission, with another three detected in the extended emission region. Like G28.30$-$0.39, the GLOSTAR-D+Eff Ly-photon rate is also greater than the total contribution from all candidate stars, which can be attributed to the non-detection of a few more candidate ionizing stars. The radio source towards the north is another H~{\small II} region located at a similar distance to G28.81$+$0.17.

\begin{table*}
\caption{Estimation of the fraction ($f_d$, $f'_d$, and $f''_d$) of Ly-photons that doesn't contribute to the ionization of hydrogen.}
\label{tab:dustfraction}
\centering
\begin{tabular}{c c c c c c c c}
\hline\hline
Source name & $\log(N^{\text{IR}}_{\text{Ly}})$  & $\log(N^{\text{RC}}_{\text{Ly}})$ & $f_d$ & $\log(N^{\text{RC}}_{\text{Ly}})$ & $f'_d$ & $\log(N^{\text{RC}}_{\text{Ly}})$  & $f''_d$ \\
 &  & (D+Eff) &  & (uGMRT) &  & (MAGPIS) &  \\
\hline
G19.68$-$0.13 & 48.91 & 48.85 & 0.12 & 48.82 & 0.19 & 48.52 & 0.59 \\
G20.99$+$0.09 & 47.69 & 47.55 & 0.28 & 47.54 & 0.29 & 47.25 & 0.64 \\
G22.76$-$0.48 & 48.69 & 48.63 & 0.13 & 48.56 & 0.26 & 47.63 & 0.91 \\
G24.47$+$0.49 & 49.14 & 49.12 & 0.05 & 48.98 & 0.30 & 48.76 & 0.58 \\
G25.69$+$0.03 & 49.10 & 49.06 & 0.09 & 49.02 & 0.17 & 48.26 & 0.85 \\
G27.49$+$0.19 & 47.98 & 47.87 & 0.21 & 47.87 & 0.22 & 47.34 & 0.77 \\
G28.30$-$0.39 & 48.92 & 48.80 & 0.23 & 48.80 & 0.24 & 48.54 & 0.58 \\
G28.81$+$0.17 & 48.97 & 48.90 & 0.15 & 48.89 & 0.17 & 48.68 & 0.48 \\
\hline
\end{tabular}
\tablefoot{RC = radio continuum, IR = infrared, $f_d$ = fraction of Ly-photons not observed in the GLOSTAR-D+Eff radio continuum emission, $f'_d$ = fraction of Ly-photons not observed in the uGMRT radio continuum emission, $f''_d$ = fraction of Ly-photons not observed in the MAGPIS radio continuum emission.}
\end{table*}

%--------------------------------------------------------------------

\section{Discussion}\label{sec:discuss}

\subsection{Nature of the extended emission}

Following the methods (described in Sect.~\ref{sec:intro}) proposed by \cite{1999ApJ...514..232K}, we have found that our target sources do not show any discontinuity in the continuum emission. Moreover, the velocity field of the RRLs shows a smooth distribution in all sources. However, as discussed in Sect.~\ref{sec:notes}, the location of the UC core in G25.69$+$0.03 with respect to the diffuse extended emission makes it unlikely to be the source of the extended emission. Hence, we conclude that the extended emission is physically associated with the compact emission for the entire sample except for G25.69$+$0.03. Thus, the gas giving rise to the diffuse emission is mostly likely to originate from the same stars responsible for the compact emission. However, a caveat in this aspect is that the velocity field is determined at a resolution of 25$\arcsec$, at which scale any discontinuity between the velocity of compact and extended emission can be difficult to ascertain (due to spatial smoothing). However, due to the weak strength of the RRLs, it is not possible to image the line emission at high resolution.

\begin{table*}
\caption{Comparison between the integrated flux densities of a few UCHRs common in the catalogs of WC89 and GLOSTAR-D survey.}
\label{tab:comparison}
\centering
\begin{tabular}{c c c c c c c}
\hline\hline
Source name & \textit{l} & \textit{b} & RA & DEC & $S^{\text{WC}}_{\text{int}}$ & $S^{\text{GLOS}}_{\text{int}}$ \\
 & (deg) & (deg) & (J2000) & (J2000) & (Jy) & (Jy) \\
\hline
G05.89--0.39& 5.885 & --0.392 & 18$^\text{h}$00$^\text{m}$30.38$^\text{s}$ & --24$\degr$04$\arcmin$00.22$\arcsec$ & 2.087 & $3.106 \pm 0.155$   \\
G08.14+0.23& 8.139 & 0.226 & 18$^\text{h}$03$^\text{m}$00.74$^\text{s}$ & --21$\degr$48$\arcmin$08.41$\arcsec$ & 0.184 & $4.875 \pm 0.244$  \\
G08.67--0.36& 8.669 & --0.356 & 18$^\text{h}$06$^\text{m}$19.02$^\text{s}$ & --21$\degr$37$\arcmin$32.35$\arcsec$ & 0.630 & $1.202 \pm 0.060$   \\
G10.62--0.38& 10.623 & --0.384 & 18$^\text{h}$10$^\text{m}$28.67$^\text{s}$ & --19$\degr$55$\arcmin$49.48$\arcsec$ & 1.207 & $4.345 \pm 0.217$  \\
G11.94--0.62& 11.937 & --0.616 & 18$^\text{h}$14$^\text{m}$01.01$^\text{s}$ & --18$\degr$53$\arcmin$25.01$\arcsec$ & 0.879 & $1.424 \pm 0.071$   \\
G19.07--0.28& 19.071 & --0.280 & 18$^\text{h}$26$^\text{m}$46.40$^\text{s}$ & --12$\degr$26$\arcmin$26.81$\arcsec$ & 0.183 & $1.352 \pm 0.068$   \\
G31.41$+$0.31& 31.413 & 0.306 & 18$^\text{h}$47$^\text{m}$34.57$^\text{s}$& --01$\degr$12$\arcmin$43.11$\arcsec$ & 0.382 & $1.080 \pm 0.054$  \\
G33.92$+$0.11& 33.915 & 0.110 &18$^\text{h}$52$^\text{m}$50.23$^\text{s}$& 00$\degr$55$\arcmin$29.35$\arcsec$ & 0.357 & $1.112 \pm 0.055$ \\
G33.50$+$0.20& 33.498 & 0.194 &18$^\text{h}$51$^\text{m}$46.72$^\text{s}$& 00$\degr$35$\arcmin$32.29$\arcsec$ & 0.736 & $0.915 \pm 0.046$ \\
G35.05$-$0.52& 35.053 & --0.518 &18$^\text{h}$57$^\text{m}$09.04$^\text{s}$& 01$\degr$39$\arcmin$03.33$\arcsec$ & 0.091 & $0.218 \pm 0.012$ \\
G41.74+0.10& 41.742 & 0.097 & 19$^\text{h}$07$^\text{m}$15.50$^\text{s}$ & +07$\degr$52$\arcmin$43.86$\arcsec$ & 0.100 & $0.323 \pm 0.016$  \\
G43.18--0.52& 43.178 & --0.518 & 19$^\text{h}$12$^\text{m}$08.68$^\text{s}$ & +08$\degr$52$\arcmin$08.72$\arcsec$ & 0.326 & $0.729 \pm 0.036$  \\
G45.12+0.13& 45.122 & 0.132 & 19$^\text{h}$13$^\text{m}$27.84$^\text{s}$ & +10$\degr$53$\arcmin$36.71$\arcsec$ & 1.167 & $5.545 \pm 0.277$  \\
G45.48+0.13& 45.478 & 0.130 & 19$^\text{h}$14$^\text{m}$08.81$^\text{s}$ & +11$\degr$12$\arcmin$27.96$\arcsec$ & 0.344 & $1.623 \pm 0.081$  \\
\hline
\end{tabular}
\tablefoot{$S^{\text{WC}}_{\text{int}}$ = integrated flux densities from WC89, $S^{\text{GLOS}}_{\text{int}}$ = integrated flux densities from the GLOSTAR-D catalog.}
\end{table*}

\subsection{Absorption of the Ly-photons by dust}

We have estimated the Ly-photon rate required to explain the radio emission using Eq.~\ref{eq:eq1}, tabulated in Table~\ref{tab:radioresults}. Since the infrared emission results from radiation from the central stars that has been reprocessed by dust, one can obtain an independent estimate of the Ly-photon rate from the infrared luminosity. We use the following equation from \citet{2001ApJ...555..613I} for this purpose:
\begin{equation}\label{eq:df}
    \left( \frac{L^{\text{\,dust}}_{\text{IR}}}{L_{\odot}} \right) \left( \frac{5.63\times10^{43}\,\text{s}^{-1}}{N_{\text{Ly}}} \right) = \frac{1-0.28f}{f},
\end{equation}
where $L^{\text{\,dust}}_{\text{IR}}$ is the integrated infrared luminosity from 8$-$1000~$\mu$m, and $f$ is the fraction of Ly-photons that contribute to the ionization of hydrogen. The values of $f_d$ ($=1-f$) for our sample are tabulated in Table~\ref{tab:dustfraction}.

As mentioned in Sect.~\ref{sec:intro}, initial interferometric studies of UCHRs showed significantly larger Ly-photon rates from the infrared compared to what was deduced from radio emission (WC89; \citealt{1994ApJS...91..659K}). Since dust is an efficient absorber of high energy photons, which in turn reduces the number of photons that can contribute to the ionization of the surrounding gas, WC89 inferred that UCHRs have significant amounts of dust that can absorb up to 99\% of the ionizing photons from the massive stars. Later studies suggested that a large fraction of the discrepancy in Ly-photon rates between radio and infrared studies was due to the presence of extended diffuse radio emission that was undetected in the early interferometer observations (e.g., \citealt{1999ApJ...514..232K, 2020MNRAS.492..895D}). Using the GLOSTAR-D+Eff data, we find that the fraction of Ly-photons that are absorbed by dust ($f_d = 1-f$) is within 5$-$28\% for our entire sample. These values are much smaller than the value of $f_d$ in WC89. This clearly highlights the role of extended emission in resolving the discrepancy in the Ly-photon rate between infrared and high angular resolution radio studies. High-resolution radio observations are only sensitive to the compact emission, while the infrared emission originates from a much larger region that encompasses the extended emission. Even in sources like G25.69$+$0.03, where the UCHR is probably not responsible for the extended emission, the Ly-photon rate inferred from infrared would be significantly overestimated, leading to an incorrect interpretation of $f_d$ to be very high.

To further examine the role of sensitivity of interferometer observations, we have used data from \textit{The Multi-Array Galactic Plane Imaging Survey} (MAGPIS; \citealt{2006AJ....131.2525H}). The MAGPIS is a radio survey of the first Galactic quadrant at 6~cm and 20~cm using the VLA in its B, C, and D-configurations. This survey has an angular resolution of $\sim$ 6$''$ and sensitivity of 1$-$2~mJy at 20~cm. We have used the 20-cm radio continuum data with VLA in its B-configuration to calculate the Ly-photon rate, which in turn is used to calculate $f''$ using Eq.~\ref{eq:df}. We have chosen the B-configuration of the VLA since the antennas are much more extended in this configuration, with much poorer sensitivity to extended emission compared to the GLOSTAR-D+Eff and uGMRT. As can be seen in Table~\ref{tab:dustfraction}, the values of $f''_d$ ($=1-f''$) are much larger than $f_d$ and $f'_d$ (going up to 91\% for G22.76$-$0.48). This conclusively establishes the role of sensitivity to extended emission at radio wavelengths in inferring the fraction of Ly-photons absorbed by dust.

Although we have demonstrated the role of extended emission in significantly reducing the need for absorption of Ly-photons by dust in our sample, one also needs to address the question of the fraction of the overall population H~{\small II} regions in the Galaxy that have compact cores surrounded by associated extended emission. While we reserve a systematic study for future work, a preliminary attempt can be made by comparing the integrated flux density of UCHRs reported by WC89 that were measured from observations at very high angular resolution ($\sim 0''.4$) with that of GLOSTAR-D which has much better sensitivity to extended emission. We find 14 UCHRs from the sample of WC89 that are present in the GLOSTAR-D survey catalog (\citealt{2019A&A...627A.175M,2024medina}). We also find two UCHRs (G23.96$-$0.15 and G41.71$+$0.11) in WC89 that are not detected in both GLOSTAR-D and THOR radio continuum maps, although their flux densities are much higher than the sensitivity of both surveys. These could be artifacts in the radio continuum maps of WC89, as we see a few bright sources near their positions.

Considering the 14 sources that are common between WC89 and GLOSTAR-D, we find that all sources have larger integrated flux densities in the GLOSTAR-D, with the ratio of flux density in GLOSTAR-D to that in WC89 ranging from 1.25 to 26.5 (see Table~\ref{tab:comparison}). This is similar to the finding of \citet{2005MNRAS.357.1003E}, who found that 8 out of 10 sources had an integrated flux density larger than 10\% when observed with the Australia Telescope Compact Array (ATCA) in the compact configuration (750D) compared to the more extended 6A configuration. Although the sample size is small, these results suggest that most H~{\small II} regions have a larger flux density at radio wavelengths than what is detected in high-resolution observations, and the presence of extended emission around these regions is one of the likely solutions to resolve the mismatch between the Ly-photon rates estimated from the radio and infrared observations. There may also be examples such as G25.69$+$0.03, where the UCHR is not directly responsible for extended emission but lies at the edge of a larger H~{\small II} region, wherein the Ly-photon rate inferred from infrared is significantly overestimated due to the infrared emission arising from the entire region.

\subsection{The ``age problem'' of UCHRs}

The number of UCHRs discovered by early radio surveys of such sources (WC89; \citealt{1994ApJS...91..659K}) was much greater than what was predicted in our Galaxy. This requires UCHRs to survive longer than their sound crossing times. This inconsistency is known as the ``age problem'' of UCHRs. One of the possible solutions to this problem may be the presence of extended emission surrounding the UC cores. The following example can illustrate this.

Following WC89, the initial radius of a Str\"{o}mgren sphere, $r_{\text{i}}$, will be $\sim 0.051$~pc for a typical UCHR (excited by an O6 star) without dust. Now, if dust absorbs 90\% (i.e., $f_d$ = 0.9) of the Ly-photons, then $r_{\text{i}}$ will be reduced by a factor of $(1 - f_d)^{1/3} = 0.46$, resulting in an initial Str\"{o}mgren radius of $\sim 0.023$~pc for $f_d$ = 0.9 (WC89). Following the balance between photoionization and recombination, the UCHR continues to expand due to the pressure difference between the UCHR and the ambient neutral medium. When the expanding ionized gas achieves pressure equilibrium, the final Str\"{o}mgren radius, $r_{\text{f}}$, reaches 
\begin{equation}
    \label{eq:expand}
    r_f = r_i \, \left(\frac{2T_e}{T_0}\right)^{2/3},
\end{equation}
where $T_{\text{e}}$ is the electron temperature of the ionized gas, and $T_0$ is the temperature of the neutral molecular gas (WC89). For a typical H~{\small II} region, taking $T_{\text{e}}$ = $10^4$~K and $T_0$ = 25~K we will get $r_{\text{f}}$ = 86~$r_{\text{i}}$. Thus, the UCHR will have a diameter of $\sim$ 2~pc at pressure equilibrium.

The time it takes for the UCHR to achieve $r_{\text{f}}$ can be estimated using the expansion rate under the strong shock approximation \citep{1980pim..book.....D},
\begin{equation}
    \label{eq:age}
    r_f (t) = r_i \, \left(1+\frac{7\,c_i\,t}{4\,r_i}\right)^{4/7},
\end{equation}
where $c_{\text{i}}$ is the sound speed ($\sim 10$~km~s$^{-1}$) inside the ionized medium. Now, following Eq.~\ref{eq:age}, the expanding UCHR will reach a size of 0.1~pc (which is the limit of a UHCR as per the classification of WC89; \citealt{1994ApJS...91..659K}) in a timescale of $1.5 \times 10^4$~years. Since the main sequence lifetime of an O6 star is $2.4 \times 10^6$~years \citep{1987A&A...182..243M}, a H~{\small II} region will be in its UC phase for roughly 0.6\% of its entire lifetime. However, the observations of \citet{2001ApJ...549..979K, 2007prpl.conf..181H, 2020MNRAS.492..895D} show a much larger number of detections of UCHRs than what is anticipated given its lifetime in the UC phase.

However, observations show that H~{\small II} regions have hierarchical density structures with compact emission co-existing with diffuse extended emission. While the exact mechanism for the formation of a H~{\small II} region with hierarchical structure is not clear (see, for example, Figure 8 of \citealt{2001ApJ...549..979K}), it is clear that simple analytical models such as that described above cannot be used to determine the age of H~{\small II} regions since the models assume the density to be uniform within the H~{\small II} region. If the extended emission is produced by leakage of ionizing photons selectively through lower-density regions within the hierarchical structure, the lifetime of the UCHR can be significantly larger than what is predicted by Eq.~\ref{eq:age} using an $r_f$ of 0.1~pc. Thus, the co-existence of extended and compact radio emission could resolve the age problem of UCHRs, although a definitive conclusion in this regard requires simulations of an expanding H~{\small II} region in a medium with a hierarchical density structure.

\section{Conclusions}

In this study, we investigate the role of extended emission surrounding H~{\small II} regions in resolving the discrepancies between the Ly-photon rates derived from the radio and infrared emission. For this purpose, we have used radio observations of eight compact and ultracompact H~{\small II} regions using the uGMRT and data from the GLOSTAR survey along with complementary infrared data from the Hi-GAL, MIPSGAL, GLIMPSE, 2MASS, and UKIDSS surveys. We have listed our main findings below.

\begin{itemize}
    \item We have detected arcmin-scale extended continuum emission surrounding all of our target H~{\small II} regions. These detections significantly boost the rate of Ly-photons at radio wavelengths compared to what would have been detected using high angular resolution observations. However, we find one example where we find the UCHR is located at the edge of the extended emission, where the Ly-photon rate may be over estimated.

    \item Besides radio continuum, we have also detected RRLs toward our targets. The RRL velocity maps show continuous velocity distributions across the target regions, also indicating a physical association between the compact core(s) and extended emission. The line widths of the RRLs (24 to 30 km~s$^{-1}$) are in agreement with what is expected for compact and ultracompact H~{\small II} regions.

    \item The in-band (GLOSTAR-D) and inter-band (uGMRT and GLOSTAR-D) spectral indices of the sources of our target sample are close to --0.1, which indicates the radio emission is optically thin. However, this is in contrast to the spectral indices of the compact emission, which is consistent with optically thick thermal emission. Thus, the extended emission dominates the properties of H~{\small II} regions at large scales.

%    \item The dust temperature of the sources of our sample ranges from 20 to 30~K, and the peak hydrogen column density ranges from $0.5 \times 10^{22}$ to $2.0 \times 10^{22}$~cm$^{-2}$. The temperatures are comparable to but slightly higher than what is found around 6.7~GHz methanol masers, and the column densities in our sample are consistent with those of other massive star-forming regions. However, they are lower than the median value, indicating that the regions are in the clearing phase of star formation.

    \item Near-infrared data from the 2MASS and UKIDSS surveys reveal multiple candidate ionizing stars within the target sources. We find that the Ly-photon rate estimated from the radio emission and the total rate from all candidate ionizing stars are within reasonable agreement. This indicates that the emissions from dense and diffuse components have originated from the same group of ionizing sources. This discovery also showcases the formation of massive stars in clusters. 

    \item The inclusion of extended emission indicates that a much smaller fraction of Ly-photons is absorbed by dust than what would be inferred from just the dense emission. In addition to reducing the estimated quantity of dust in H~{\small II} regions, the presence of extended emission may resolve the ``age problem'' of the UCHRs.
\end{itemize}
The results of our work highlight the importance of sensitivity to detect extended emission, which is one of the key highlights of the full GLOSTAR survey.

%--------------------------------------------------------------------

\begin{acknowledgements}
The authors thank the referee for a critical review, which helped to improve the clarity of the paper. J.D. and J.D.P. thank the Max Planck Society for funding this research through the Max Planck Partner Group. D.V.L. acknowledges the support of the Department of Atomic Energy, Government of India, under project no. 12-R\&D-TFR-5.02-0700. S.A.D. acknowledges the M2FINDERS project from the European Research Council (ERC) under the European Union's Horizon 2020 research and innovation
programme (grant No 101018682). We thank the staff of the uGMRT that made these observations possible. uGMRT is run by the National Centre for Radio Astrophysics of the Tata Institute of Fundamental Research. The VLA is run by the National Radio Astronomy Observatory, a facility of the National Science Foundation operated under a cooperative agreement by Associated Universities, Inc. This work (partially) uses information from the GLOSTAR database at \url{http://glostar.mpifr-bonn.mpg.de} supported by the MPIfR, Bonn. J.D. also thanks Dr.~Jean-Baptiste~Jolly from the Max Planck Institute for Extraterrestrial Physics and the Nordic ALMA Regional Center node for providing the \texttt{Line-Stacker} package. This research has made use of \texttt{Photutils}, an \texttt{Astropy} package for the detection and photometry of astronomical sources. Additionally, the SIMBAD database, operated at CDS, Strasbourg, France, has been utilized. This research has also made use of resources such as NASA's Astrophysics Data System, CDS's VizieR catalog access tool, and calibrated infrared images from the Spitzer Heritage Archive and Herschel Science Archive.
\end{acknowledgements}

%--------------------------------------------------------------------

\bibliographystyle{aa} % style aa.bst
\bibliography{reference.bib} % your references Yourfile.bib

%--------------------------------------------------------------------

\appendix

\section{Estimation of the electron temperature}
\label{appendix}

Under the local thermodynamic equilibrium (LTE) conditions and Rayleigh-Jeans limit, the optical depth for the continuum emission ($\tau_{\text{C}}(\nu)$) can be derived as,

\begin{equation}
  \label{eq:app1}
  \tau_{\text{C}}(\nu) = - \ln \left( 1 - \frac{{c}^2\,{I}_{\text{C}}(\nu)}{2\,{k}_{\text{B}}\,{T}_{\text{e}}\,\nu^2} \right),
\end{equation}
where $\nu$ is the frequency, $k_{\text{B}}$ is the Boltzmann constant, $T_{\text{e}}$ is the electron temperature, and $I_{\text{C}}$ is the specific intensity of the continuum emission. Similarly, the optical depth for the line emission ($\tau_{\text{L}}(\nu)$) can be derived as,

\begin{equation}
    \label{eq:app2}
    \tau_{\text{L}}(\nu) = - \ln \left( 1-\frac{{I}_{\text{L}}}{{I}_{\text{C}}}\, {e}^{\tau_{\text{C}}(\nu)}\, \left[1-{e}^{-\tau_{\text{C}}(\nu)}\right] \right),
\end{equation}
where $I_{\text{L}}$ is the  specific intensity of the line emission. Using Eq.~\ref{eq:app1} and Eq.~\ref{eq:app2}, one can determine the total optical depths of continuum and line emissions by measuring the total $I_{\text{C}}$ and $I_{\text{L}}$, respectively. The $\tau_{\text{C}}(\nu)$ along
a line of sight (LOS) can also be related to the properties of the H~{\small II} region using the Altenhoff approximation \citep{1961VeBon..59....1A}, 

\begin{equation}
\label{eq:app3}
    \tau_{\text{C}}(\nu) \approx 8.235 \times 10^{-2} \,\left(\frac{{T}_{\text{e}}}{\text{K}}\right)^{-1.35} \,\left(\frac{\nu}{\text{GHz}}\right)^{-2.1}\, \left(\frac{EM}{\text{pc}.\text{cm}^{-6}}\right),
\end{equation}
where $EM$ is the emission measure. The optical depth $\tau_{\text{L}}(\nu)$ corresponds to the center frequency of the line in terms of the emission measure given by the following \citep{2013tra..book.....W},

\begin{equation}
\label{eq:app4}
    \tau_{\text{L}}(\nu) \approx 1.92 \times 10^{3} \,\left(\frac{{T}_{\text{e}}}{\text{K}}\right)^{-2.5} \,\left(\frac{\Delta\nu}{\text{KHz}}\right)^{-1}\, \left(\frac{EM}{\text{pc}.\text{cm}^{-6}}\right),
\end{equation}
where $\Delta\nu$ is the FWHM of the RRLs in KHz.

Following Eq.~\ref{eq:app3} and Eq.~\ref{eq:app4}, the
electron temperature ($T_{\text{e}}$) can be obtained as,

\begin{equation}
\label{eq:app5}
     \left(\frac{{T}_{\text{e}}}{\text{K}}\right) = \left[ 23.315 \times 10^{3} \, \frac{\tau_{\text{C}}(\nu)}{\tau_{\text{L}}(\nu)} \, \left(\frac{\Delta\nu}{\text{KHz}}\right)^{-1} \, \left(\frac{\nu}{\text{GHz}}\right)^{2.1} \right]^{1/1.15}.
\end{equation}

It is to be noted that Eq.~\ref{eq:app1} is not completely independent, and we have to provide a $T_{\text{e}}$ to derive a $\tau_{\text{C}}(\nu)$ value. When the continuum emission is optically thin, the $\tau_{\text{C}}(\nu)$ is inversely proportional to $T_{\text{e}}$, and $T_{\text{e}}$ can be derived by using Eq.~\ref{eq:app2} and Eq.~\ref{eq:app5}.

However, for moderate optical depth, the relation between the $\tau_{\text{C}}(\nu)$ and $T_{\text{e}}$ is non-linear. Thus, we adopt an iterative procedure, where an
initial guess of the $T_{\text{e}}$ is provided in
Eq.~\ref{eq:app1} following which the $T_{\text{e}}$ is recomputed using Eq.~\ref{eq:app5}. This is repeated until the value of $T_{\text{e}}$ converges.

Moreover, \citet{1966ApJ...144.1225G} showed that the observed RRL intensities could only be explained using non-LTE conditions, similar to the anomalous intensities of optical lines from nebulae and stellar atmospheres. Following \citet{1966ApJ...144.1225G}, the excitation temperature ($T_{\text{ex}}$) of an electronic transition is not equal to the $T_{\text{e}}$ of the ionized gas inside a H~{\small II} region, and is thus corrected using

\begin{equation}
    \label{eq:app6}
    {e}^{-\,{h}\,\nu/{k}_{\text{B}} {T}_{\text{ex}}} = 
    \frac{{b}_{\text{n}}}{{b}_{\text{n-1}}}\,\,{e}^{-\,{h}\,\nu/{k}_{\text{B}} {T}_{\text{e}}},
\end{equation}
where $b_{\text{n}}$, the departure coefficient that is the ratio of the actual population of atoms in the n-th state to the population that would be there if the ionized gas were in the LTE at the temperature $T_{\text{e}}$. Under the condition $h\nu << k_{\text{B}}T_{\text{e}}$, the corrected or non-LTE value of the line absorption coefficient ($\kappa_{\text{L}}$) for the line n $\rightarrow$ n--1 is found to be \citep{1966ApJ...144.1225G, 1972MNRAS.157..179B},

\begin{equation}
    \label{eq:app7}
    \kappa_{\text{L}} = \kappa^*_{\text{L}}\,{b}_{\text{n-1}}\,\beta,
\end{equation}
where $\kappa^*_{\text{L}}$ is the LTE value of the line absorption coefficient, and the correction factor ($\beta$) is approximated by \citet{1972MNRAS.157..179B} for RRL transitions with $\Delta{n} \approx 1$. If the values of $|\tau_{\text{L}}|$ and $\tau_{\text{C}}$ are less than 1, we can approximate the exponential terms in Eq.~\ref{eq:app2} to the second order to get

\begin{equation}
    \label{eq:app8}
    {I}_{\text{L}} \, \approx \, {I}^{*}_{\text{L}}\,{b}_{\text{n}}\,\left(1-\frac{\tau_{\text{C}}}{2}\beta\right),
\end{equation}
under the condition $h\nu <<k_{\text{B}}T_{\text{e}}$. Again, ${I}^*_{\text{L}}$ is the LTE value of the line temperature. Now, using the non-LTE value i.e. ${I}_{\text{L}}$ in Eq.~\ref{eq:app2} and following Eq.~\ref{eq:app5}, we can get the non-LTE electron temperature of the ionized gas.

%--------------------------------------------------------------------

\section{Figures}
\label{appendix1}

\begin{figure*}
\centering
\includegraphics[width=0.47\textwidth]{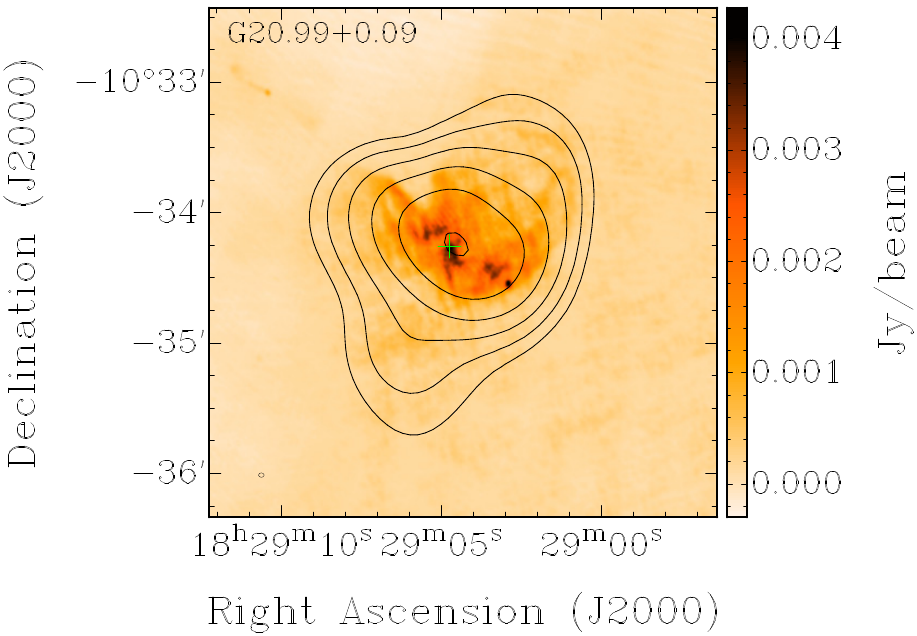}
\hspace{0.5cm}
\includegraphics[width=0.43\textwidth]{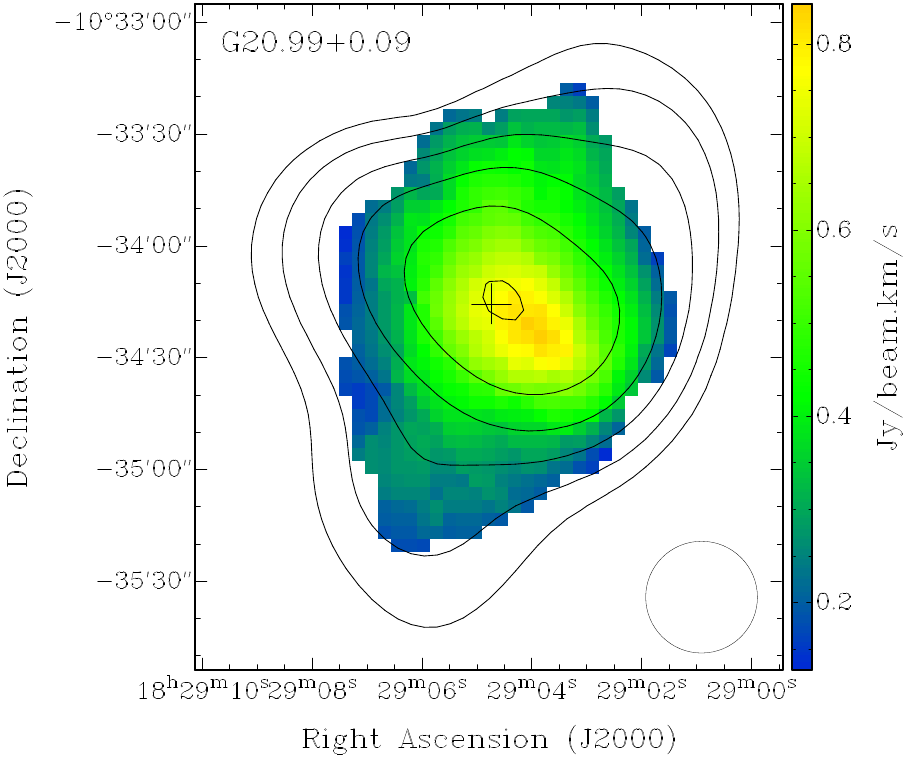}
          
\vspace{1cm}

\includegraphics[width=0.45\textwidth]{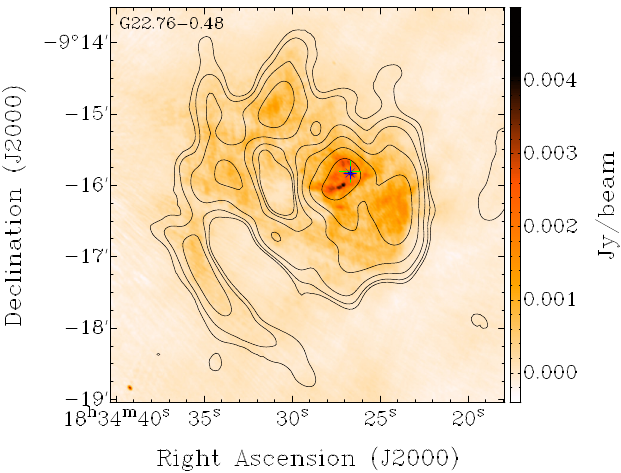}
\hspace{0.5cm}
\includegraphics[width=0.45\textwidth]{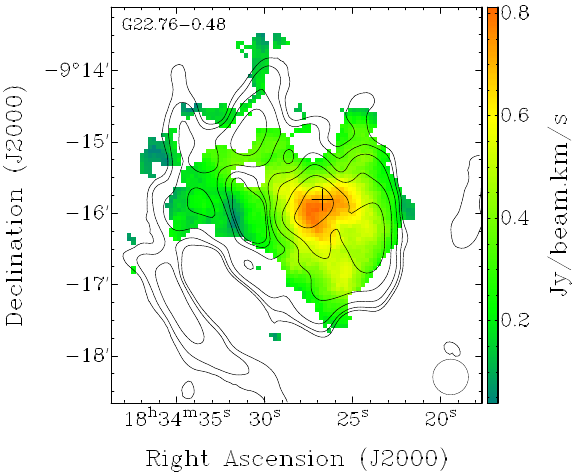}

\vspace{1cm}

\includegraphics[width=0.45\textwidth]{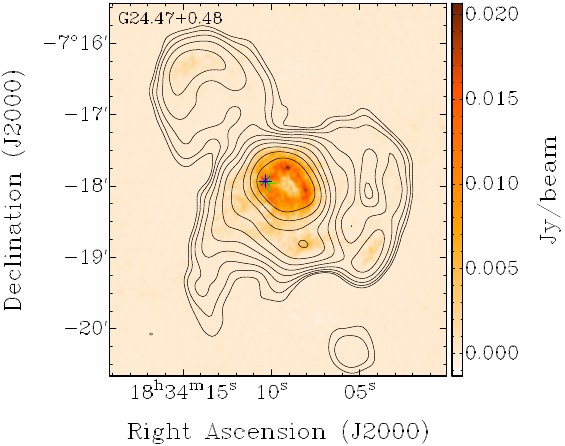}
\hspace{0.5cm}
\includegraphics[width=0.45\textwidth]{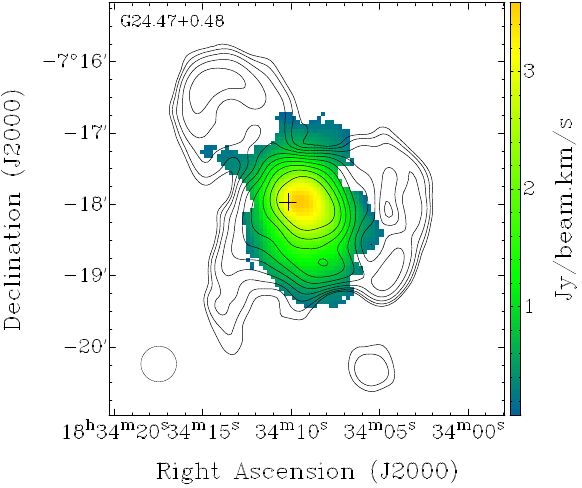}

\caption{uGMRT radio continuum (left) and GLOSTAR-D moment-0 (right) maps of individual H~{\small II} regions overlaid with the radio contours from the GLOSTAR-D continuum maps in black. The contours have started at the 3$\sigma$-level flux and have increased in steps of $\sqrt{3}$ (top), $2$ (middle), and $\sqrt{3}$ (bottom), respectively. The coordinates reported in the THOR radio continuum catalog are shown using the green/black `$+$' signs, while the locations of the UC cores (from the GLOSTAR-B and CORNISH surveys) are shown using the blue `$\ast$' signs. The respective beam sizes are shown at the bottom-left/bottom-right corners of the figures.}
\label{fig:radioemission-app}
\end{figure*}

\begin{figure*}
\addtocounter{figure}{-1}
\centering
\includegraphics[width=0.45\textwidth]{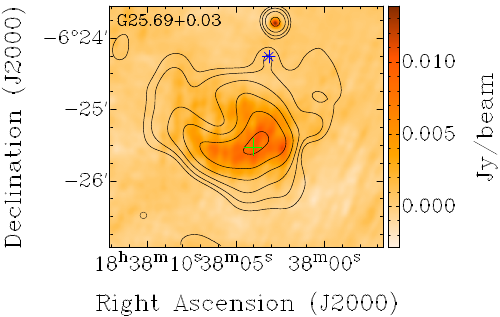}
\hfill
\includegraphics[width=0.45\textwidth]{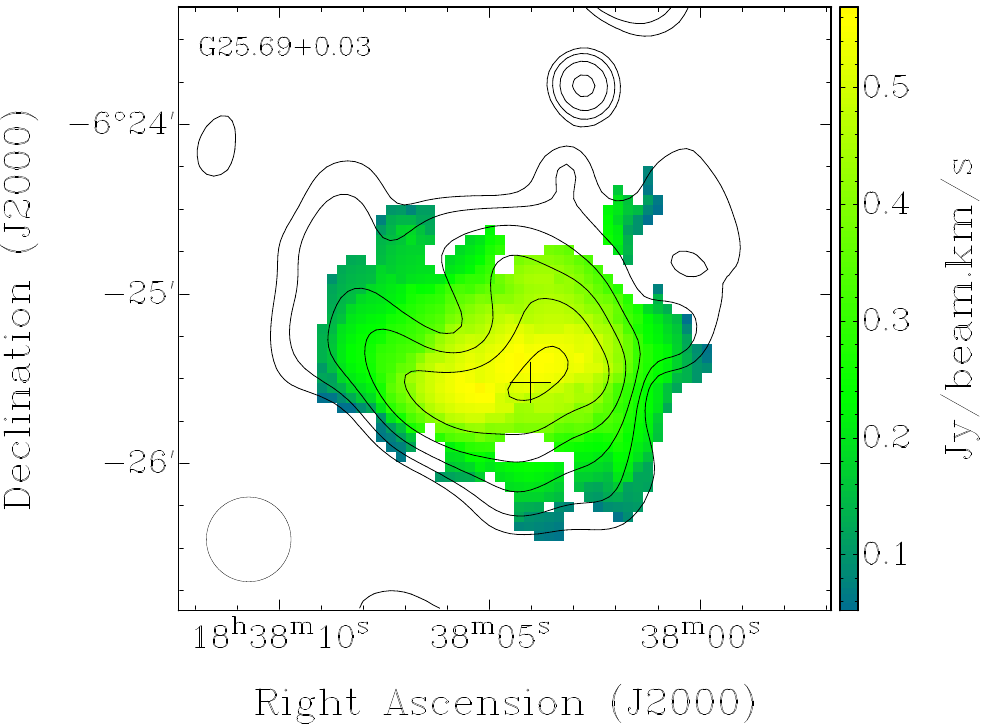}

\vspace{1cm}

\includegraphics[width=0.45\textwidth]{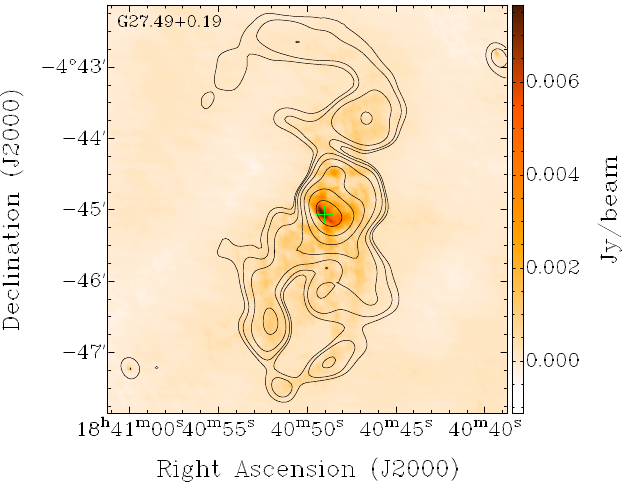}
\hfill
\includegraphics[width=0.45\textwidth]{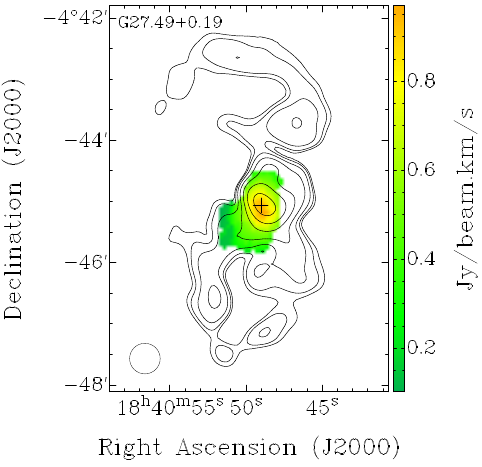}

\vspace{1cm}

\includegraphics[width=0.45\textwidth]{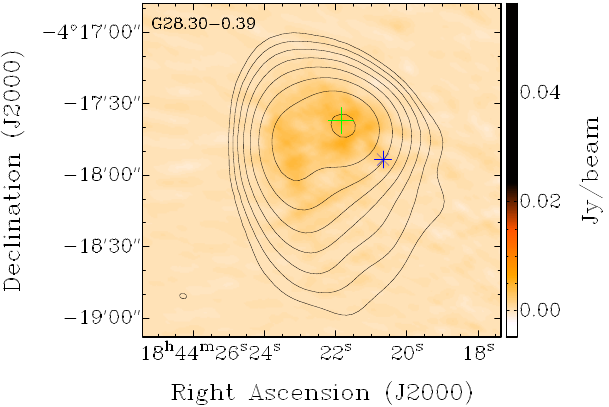}
\hfill
\includegraphics[width=0.45\textwidth]{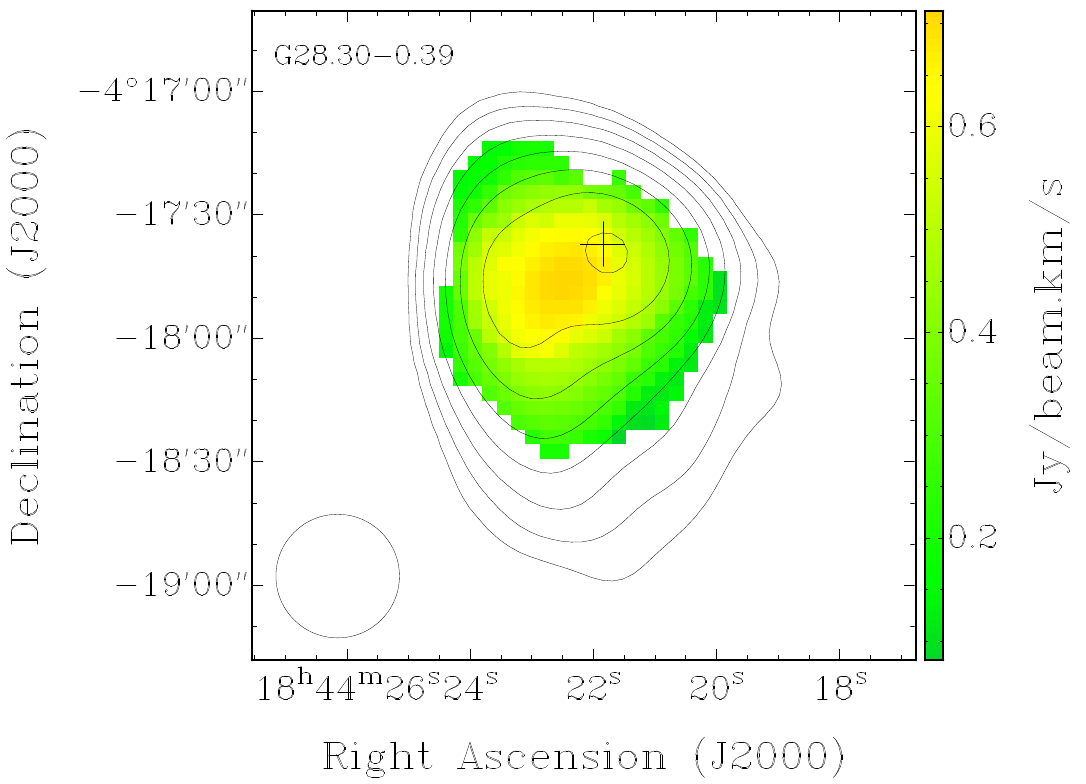}
          
\caption{Same as before, but the contours have started at the 3$\sigma$-level flux and have increased in steps of $\sqrt{3}$ (top), $2$ (middle), and $\sqrt{3}$ (bottom), respectively.}
\label{fig:radioemission-app}
\end{figure*}

\clearpage

\begin{figure*}
\addtocounter{figure}{-1}
\centering
\includegraphics[width=0.45\textwidth]{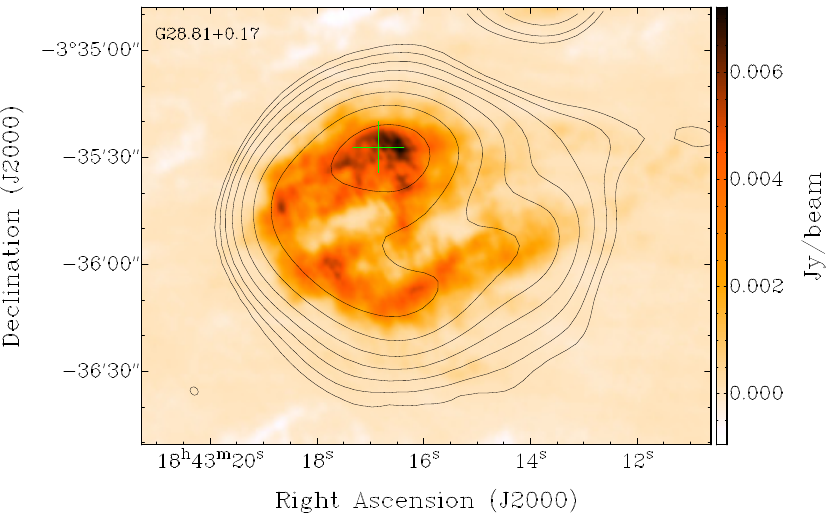}
\hfill
\includegraphics[width=0.45\textwidth]{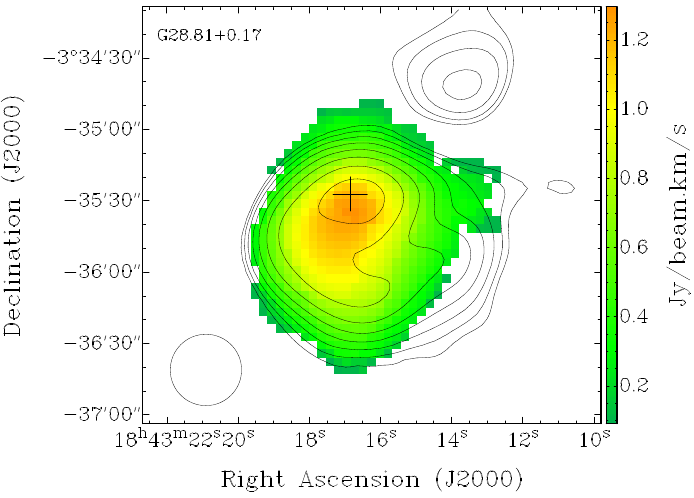}
                    
\caption{Same as before but the contours have started at the 3$\sigma$-level flux and have increased in steps of $\sqrt{3}$.}
\label{fig:radioemission-app}
\end{figure*}

\begin{figure*}
\centering
\includegraphics[width=0.45\textwidth]{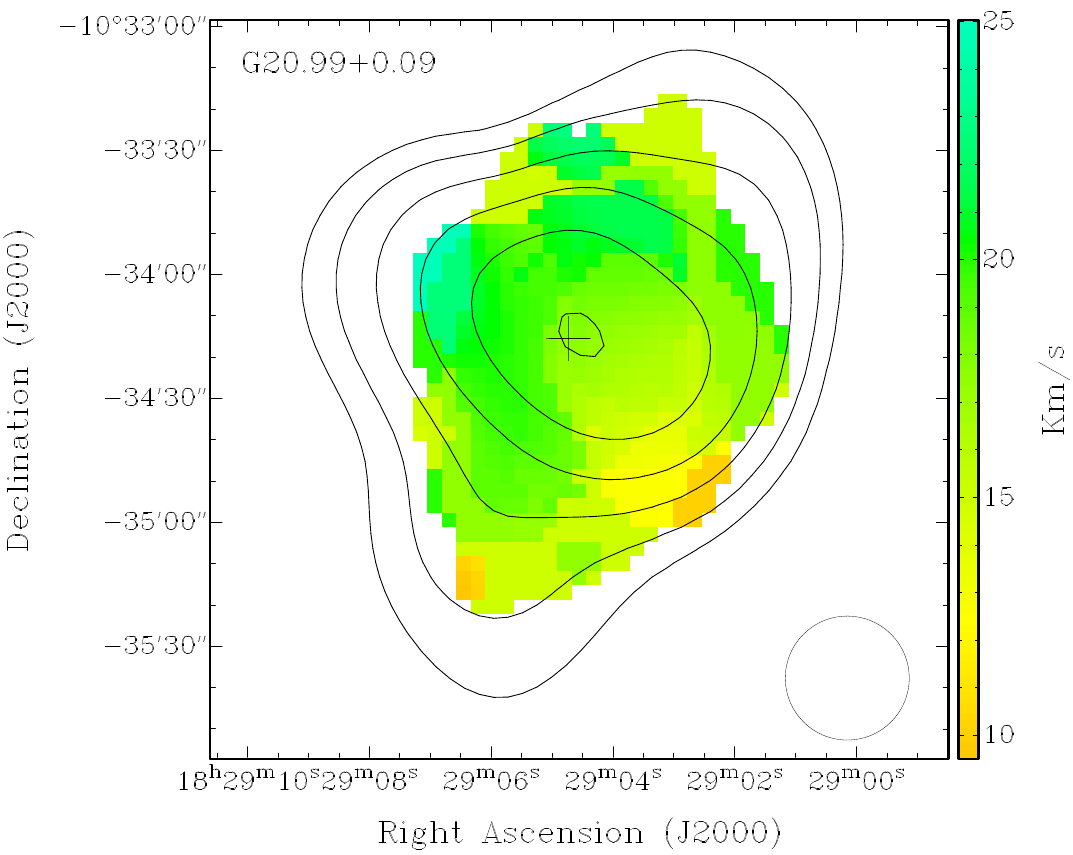}

\vspace{1cm}
          
\includegraphics[width=0.44\textwidth]{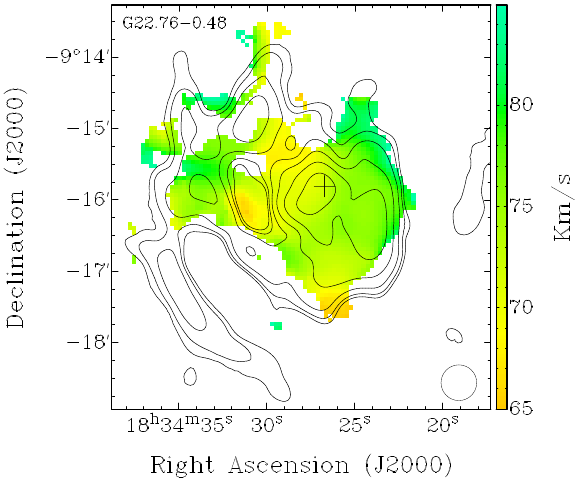}
\hfill
\includegraphics[width=0.44\textwidth]{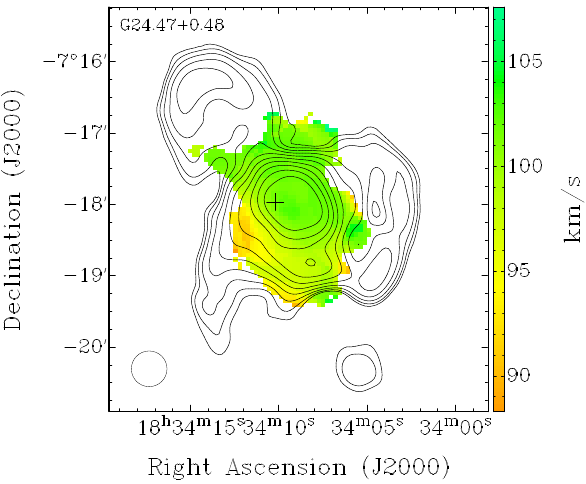}

\vspace{1cm}

\includegraphics[width=0.43\textwidth]{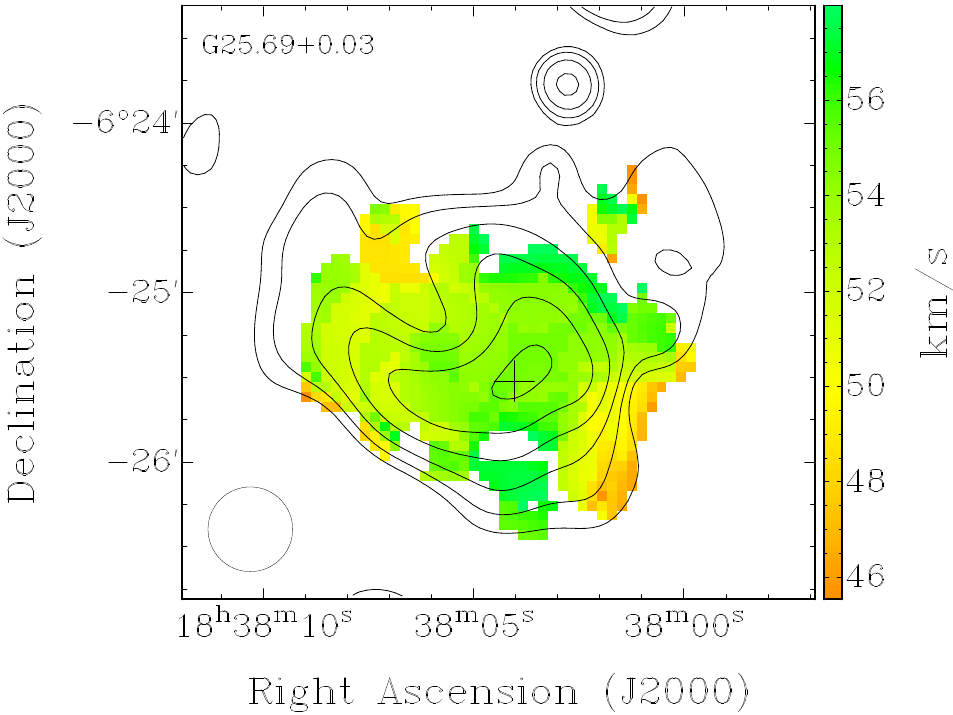}
\hfill
\includegraphics[width=0.43\textwidth]{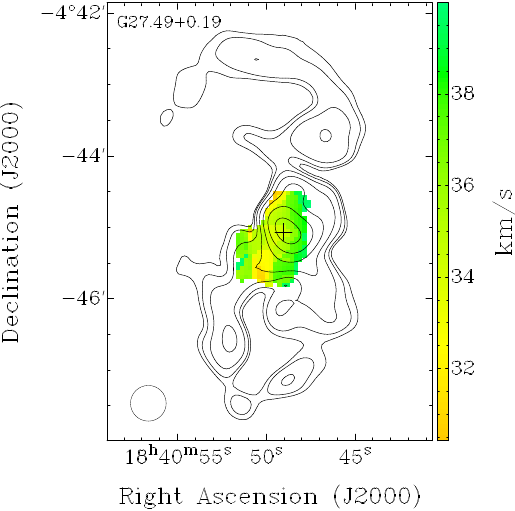}
          
\caption{RRL peak velocity distribution maps from the GLOSTAR-D data overlaid with the GLOSTAR-D radio continuum contours in black. Starting from the 3$\sigma$-level, the radio contours increase in the multiplicative steps of $\sqrt{3}$ (top), $2$ (middle left), $\sqrt{3}$ (middle right), $\sqrt{3}$ (bottom left), and $2$ (bottom right), respectively. The coordinates reported in the THOR radio continuum catalog are shown using black `$+$' signs, and the respective beam sizes are shown at the bottom-left/bottom-right corner of the figures.}
\label{fig:velocityfield-app}
\end{figure*}

\begin{figure*}
\addtocounter{figure}{-1}
\centering
\includegraphics[width=0.45\textwidth]{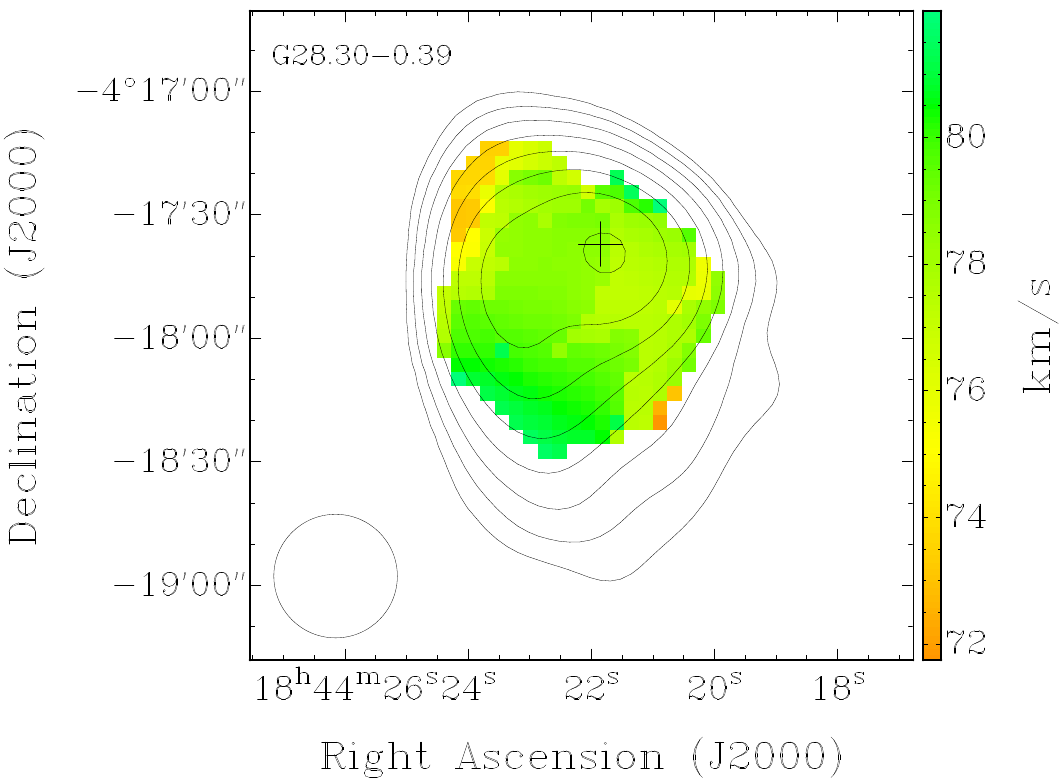}
\hfill
\includegraphics[width=0.45\textwidth]{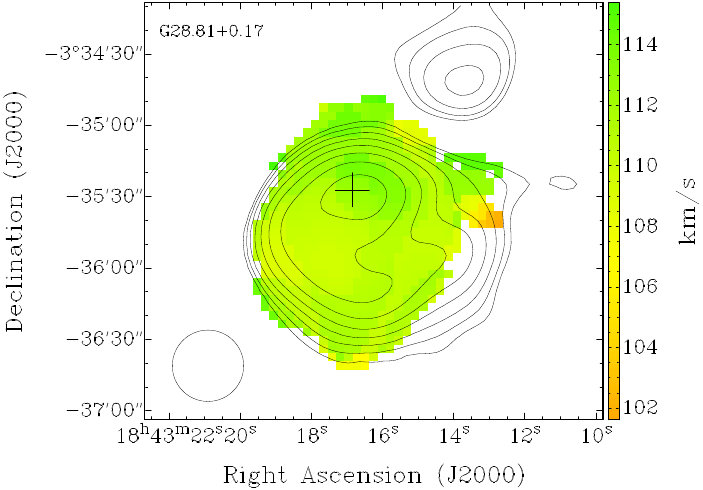}

\caption{Same as before, but the radio contours have started at the 3$\sigma$-level flux and have increased in steps of $\sqrt{3}$ (left), and $\sqrt{3}$ (right), respectively.}
\label{fig:velocityfield-app}
\end{figure*}

\begin{figure*}
\centering
\includegraphics[width=0.45\textwidth]{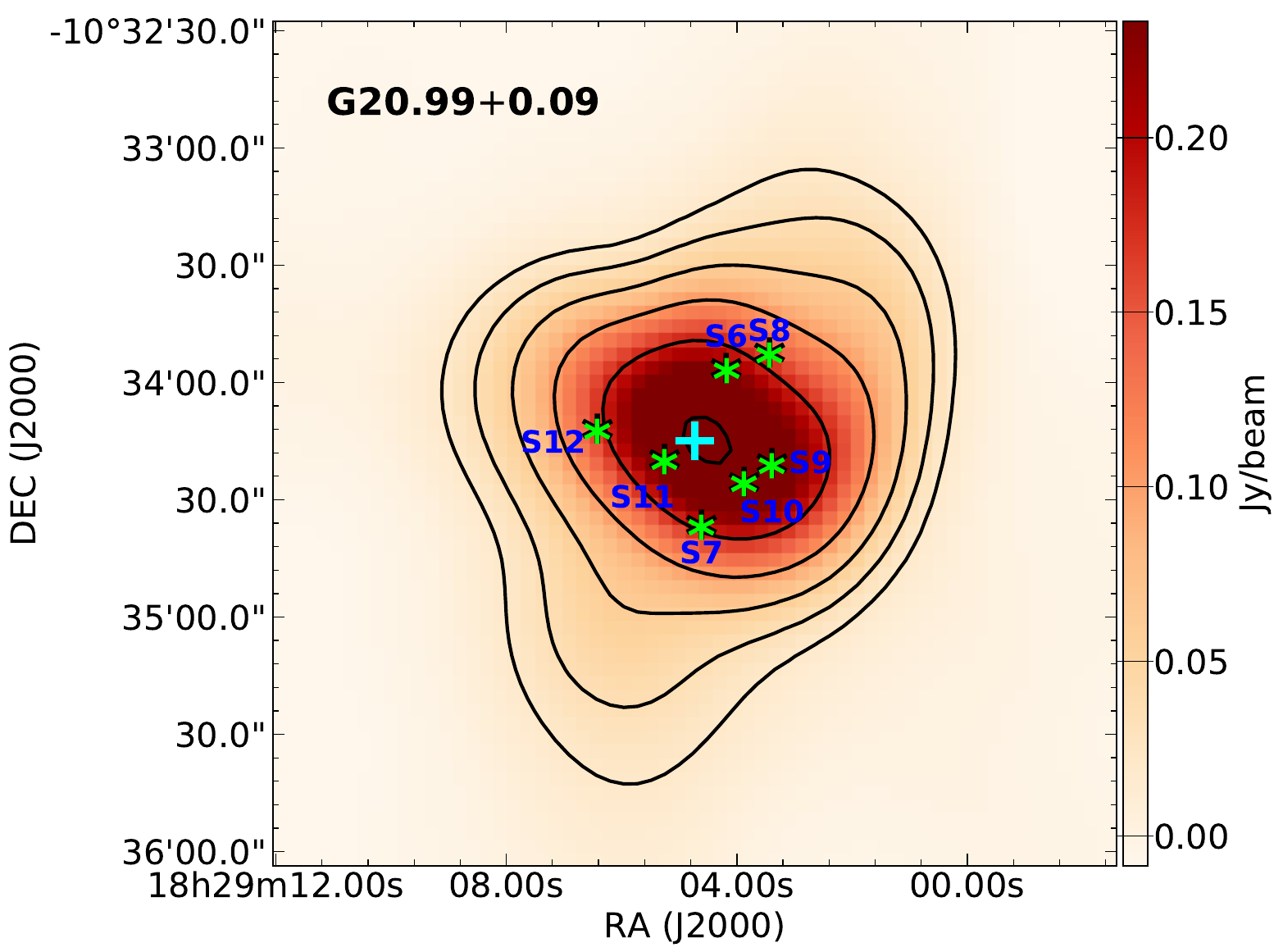}

\vspace{1cm}
        
\includegraphics[width=0.45\textwidth]{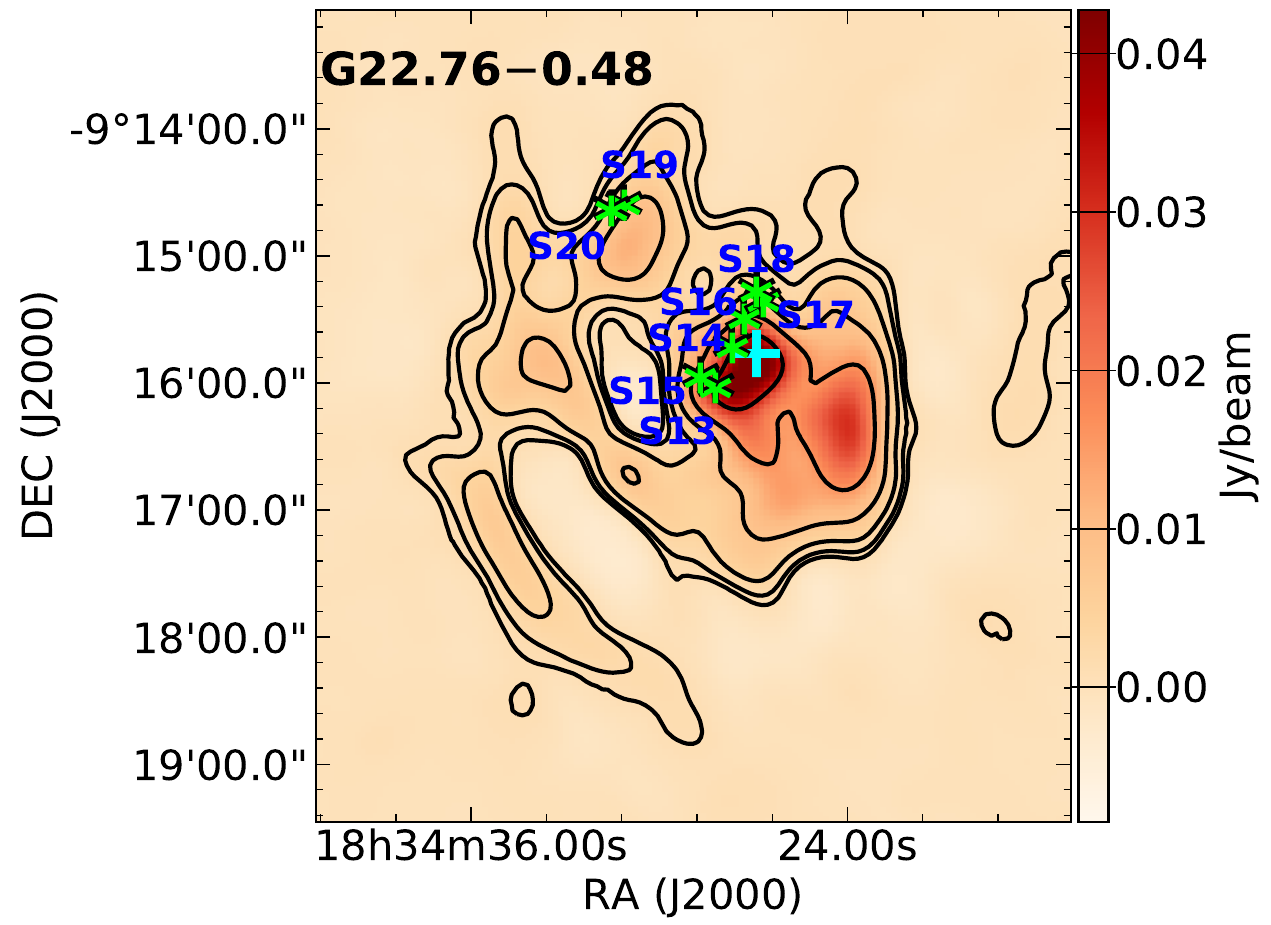}
\hfill
\includegraphics[width=0.45\textwidth]{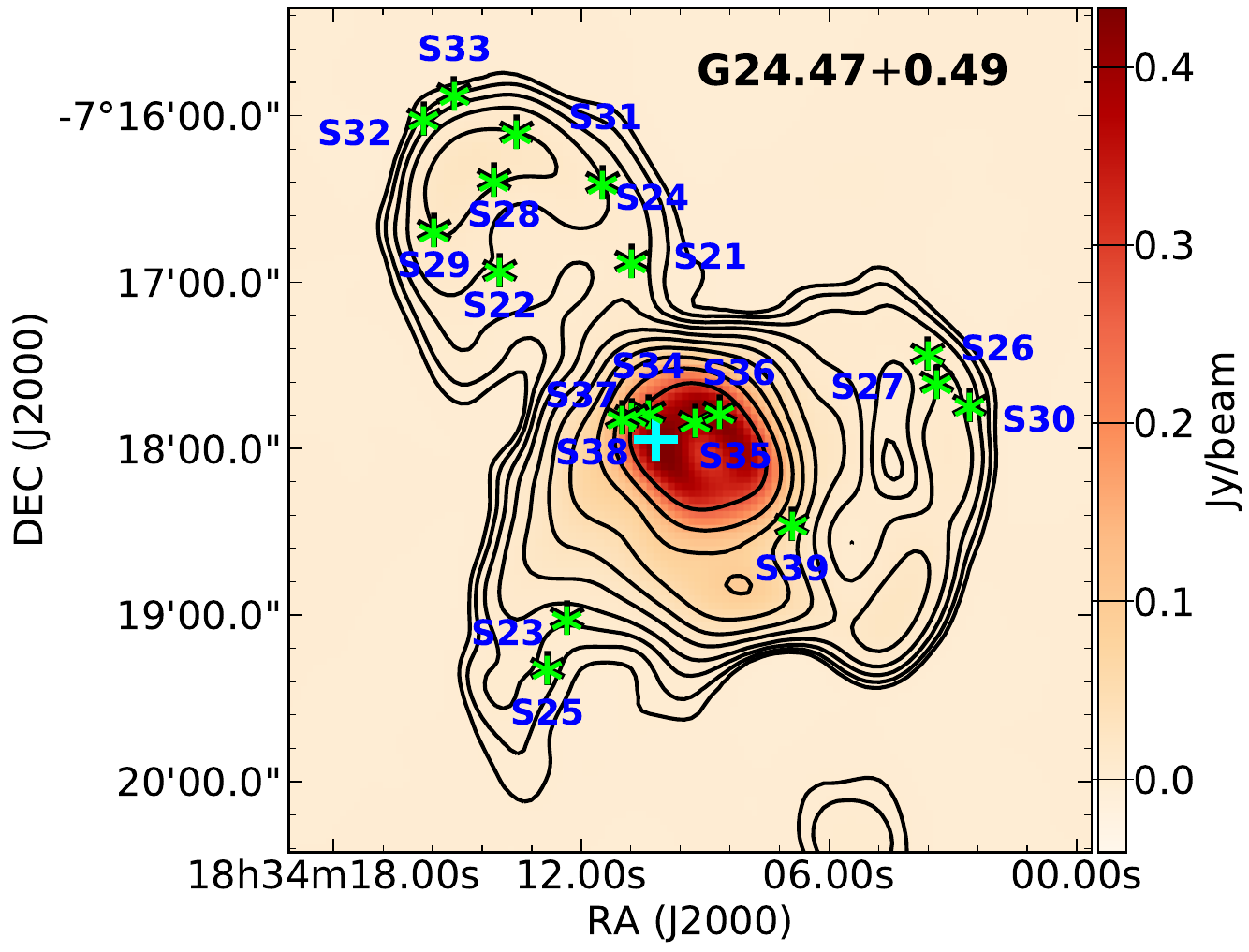}

\vspace{1cm}
          
\includegraphics[width=0.45\textwidth]{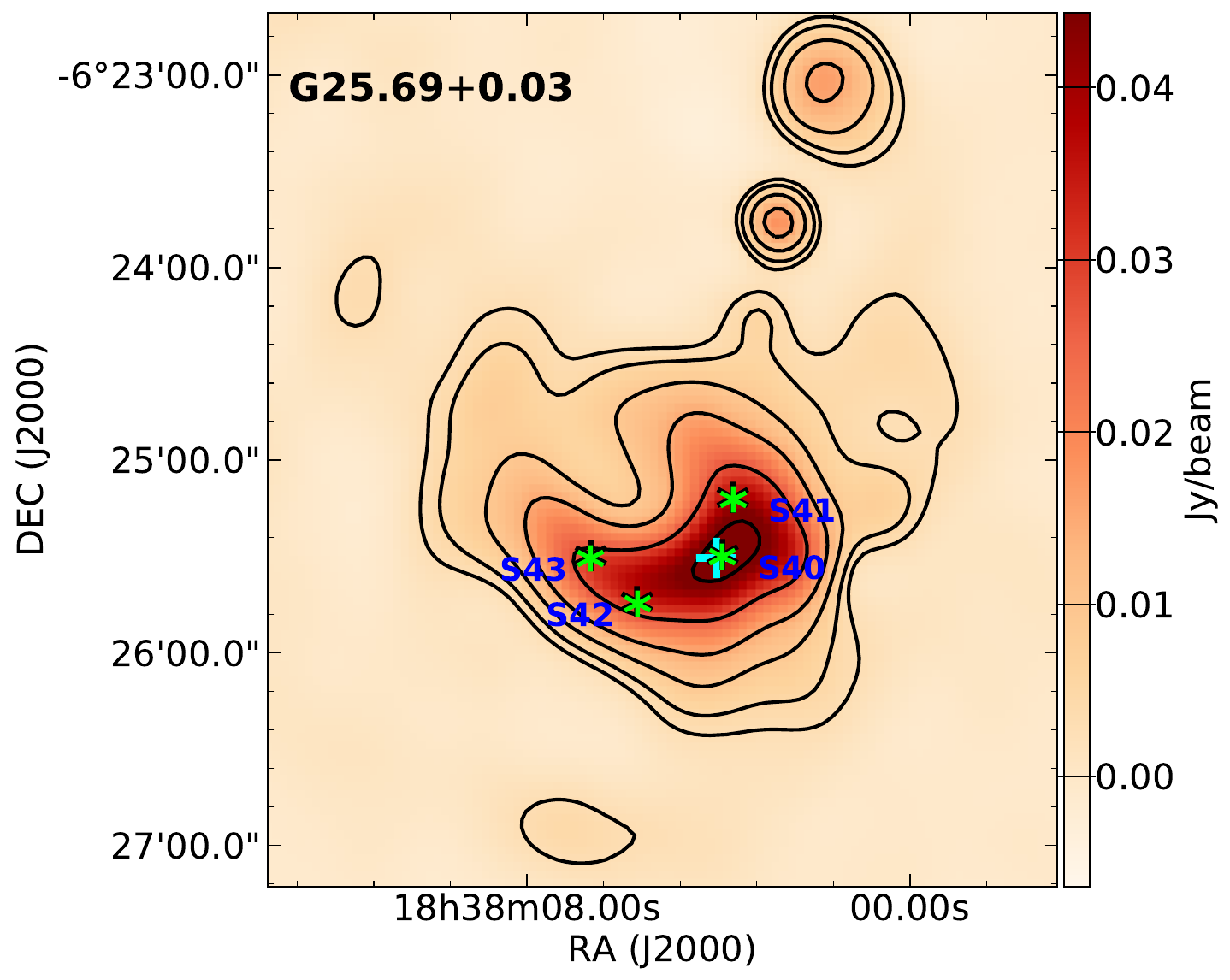}
\hfill
\includegraphics[width=0.45\textwidth]{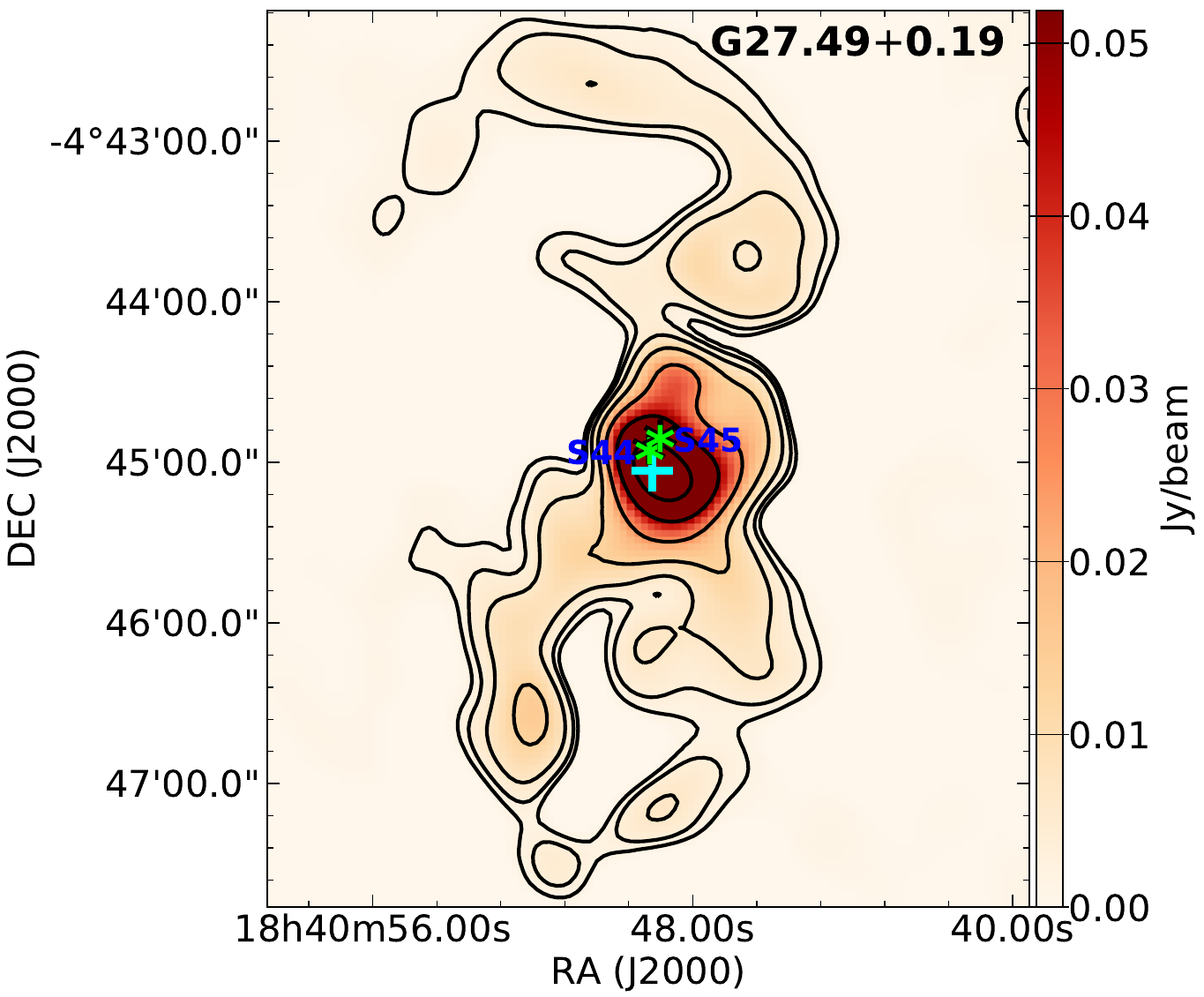}

\caption{Locations of the candidate ionizing stars (labeled using an S\_\_ format) are shown using green `*' signs for the individual H~{\small II} regions (see Table~\ref{tab:stars} for details). The cyan `+' signs indicate the coordinates reported in the THOR radio continuum catalog.
}\label{fig:starlocations-app}
\end{figure*}

\begin{figure*}
\addtocounter{figure}{-1}
\centering
\includegraphics[width=0.45\textwidth]{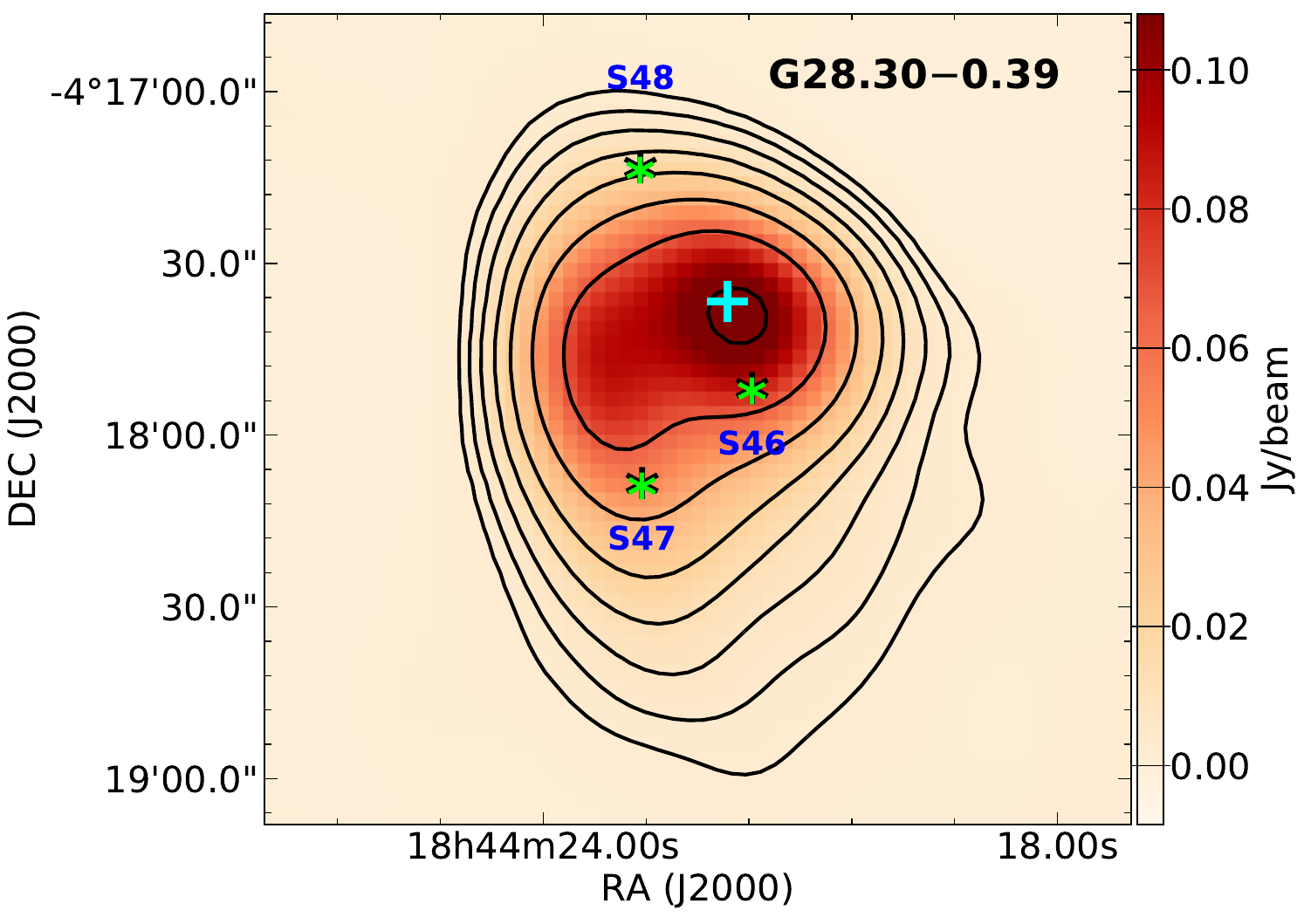}
\hfill
\includegraphics[width=0.45\textwidth]{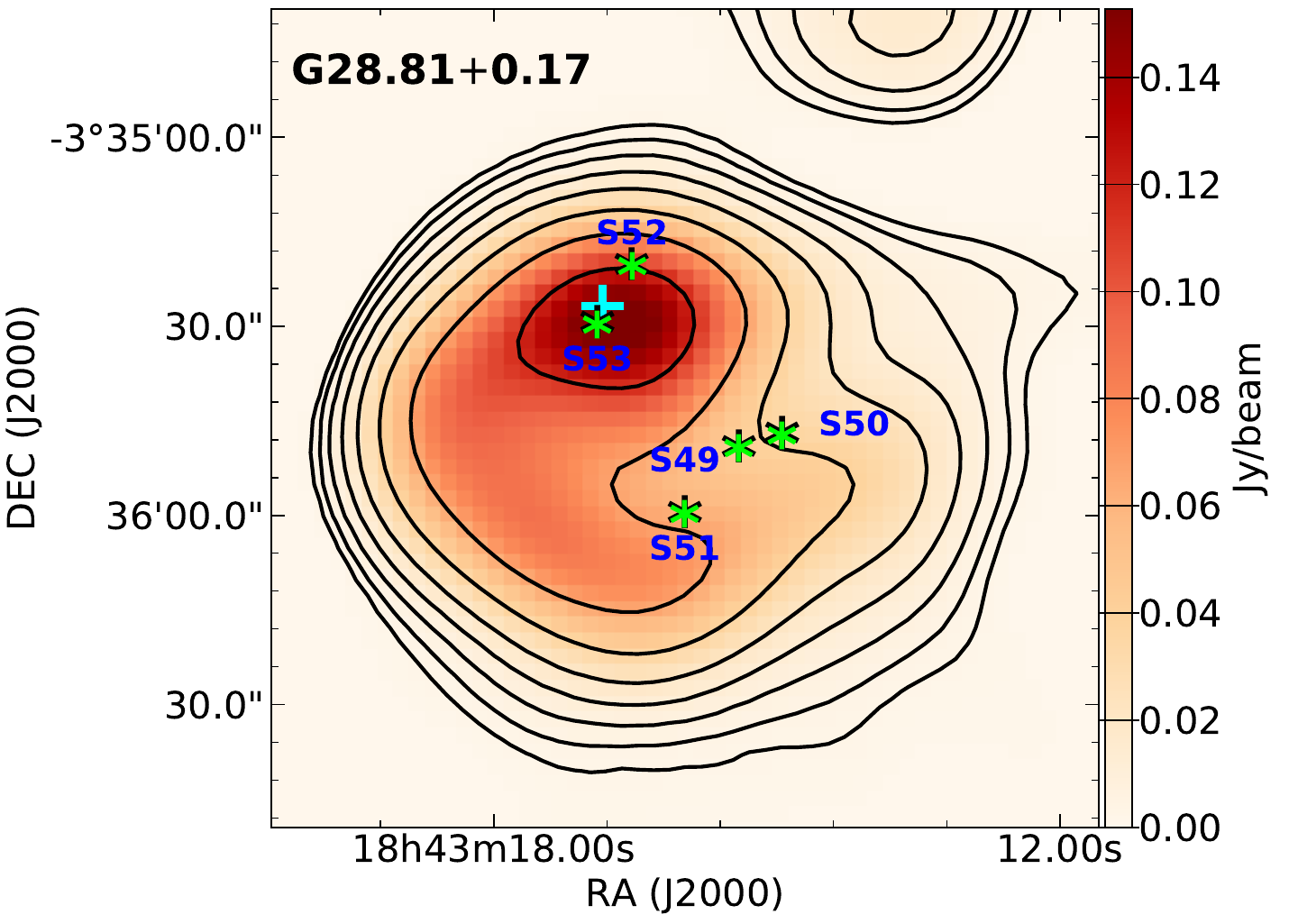}

\caption{Continued}
\label{fig:starlocations-app}
\end{figure*}

%\FloatBarrier

\end{document}